REVIEW ARTICLE

*Journal of Software: Evolution and Process* WILEY

# Understanding software architecture erosion: A systematic mapping study

Ruiyin Li[1,2] | Peng Liang[1] | Mohamed Soliman[2] | Paris Avgeriou[2]

[1]School of Computer Science, Wuhan University, Wuhan, China

[2]Department of Mathematics and Computing Science, University of Groningen, Groningen, The Netherlands

**Correspondence**
Peng Liang, School of Computer Science, Wuhan University, Wuhan, China.
Email: liangp@whu.edu.cn

**Funding information**
National Natural Science Foundation of China, Grant/Award Number: 62172311; ITEA3 and RVO, Grant/Award Number: 17038; China Scholarship Council

**Abstract**

Architecture erosion (AEr) can adversely affect software development and has received significant attention in the last decade. However, there is an absence of a comprehensive understanding of the state of research about the reasons and consequences of AEr, and the countermeasures to address AEr. This work aims at systematically investigating, identifying, and analyzing the reasons, consequences, and ways of detecting and handling AEr. With 73 studies included, the main results are as follows: (1) AEr manifests not only through architectural violations and structural issues but also causing problems in software quality and during software evolution; (2) non-technical reasons that cause AEr should receive the same attention as technical reasons, and practitioners should raise awareness of the grave consequences of AEr, thereby taking actions to tackle AEr-related issues; (3) a spectrum of approaches, tools, and measures has been proposed and employed to detect and tackle AEr; and (4) three categories of difficulties and five categories of lessons learned on tackling AEr were identified. The results can provide researchers a comprehensive understanding of AEr and help practitioners handle AEr and improve the sustainability of their architecture. More empirical studies are required to investigate the practices of detecting and addressing AEr in industrial settings.

## 1 | INTRODUCTION

Architecture may exhibit an eroding tendency when changes are accumulated in the software system. As the system evolves, the accumulation of such problems (e.g., architectural violations) can cause the implemented architecture to deviate away from the intended architecture. The phenomenon of divergence between the intended and implemented architectures is regarded as architecture erosion (AEr).[1] An eroded architecture can aggravate the brittleness of the system[1] and decrease architecture sustainability.[2] For instance, a software system with an eroded architecture may lead to the deterioration of the engineering quality of the system[3] and make it difficult for developers to understand the internal structure of the system.[1] Furthermore, AEr might make it very hard to implement new requirements and consequently negatively affect the extensibility of the system.[4]

Due to the severe consequences of AEr, it has been the subject in architecture research since the architecture concept was coined, and the first definition of the concept of AEr was given by Perry and Wolf[1] almost 30 years ago. However, the phenomenon of AEr has been described using various terms and definitions in the literature, including software erosion,[3,5] design erosion,[6] design decay,[7] architecture degeneration,[8] architectural decay,[9] code decay,[10] and modular deterioration.[11] Some of these terms focus on erosion at different levels of abstraction; for example, code decay[10] highlights more code anomalies (i.e., at the code level), whereas modular deterioration[11] concentrates on the modularity of systems (i.e., at the architecture level). These terms indicate that AEr might have an impact on different levels of software systems, and the viewpoints of the aforementioned studies also imply that AEr is a multifaceted phenomenon.

Because the research landscape on this field is diverse and perplexing, there is a need for a comprehensive overview and systematic analysis of the literature on AEr. To this end, we conducted a Systematic Mapping Study (SMS),[12] to investigate the definitions behind the AEr phenomenon, the reasons that incur it, the consequences it imposes, and the research approaches for detecting and handling AEr. This SMS aims at consolidating the existing research results and deriving research directions for future work.

Both SMS and Systematic Literature Review (SLR) are typically employed to survey the literature on a specific topic area. An SLR focuses on investigating, evaluating, and interpreting the available studies related to specific research questions towards a topic or phenomenon,[13] whereas an SMS provides an overview of a research area to systematically identify and evaluate the evidence in literature.[12] One of the main differences between an SMS and an SLR is that an SMS aims to discover the research trends and covers a broad topic in the literature, whereas an SLR usually has a relatively narrow and in-depth research scope and focuses on specific research questions.[12] Specifically, the AEr phenomenon has an impact on different levels of software systems and affects various aspects in software development (e.g., development activities).[14] Hence, to establish an overview of this topic area in the context of software development, we decided to conduct an SMS rather than an SLR on the studied topic (i.e., architecture erosion in software development).

The remainder of this paper is organized as follows: Section 2 introduces the context of architecture erosion. Section 3 elaborates on the mapping study design and the research questions. Section 4 presents the results of each research question. In Section 5, we discussed the results of the research questions and implications to researchers and practitioners. Section 6 examines the threats to validity. Section 8 summarizes this SMS.

## 2 | CONTEXT

In this section, we present the context of this SMS by briefly introducing the terms and some characteristics of AEr.

### 2.1 | Terms of architecture erosion

As mentioned in Section 1, many studies explored the phenomenon of AEr, but they described this phenomenon with different terms. In the seminal paper on software architecture, Perry and Wolf[1] coined the concept of AEr and drift and they argued that AEr happens due to the violations of architecture. "Architecture decay" is also a common term used to describe this phenomenon; for example, Behnamghader et al.[15] explored AEr through a large-scale empirical study of architectural evolution in open-source projects. Dalgarno[5] used "software erosion" to denote the constant internal structural decay of software systems, which manifests that the implemented architecture diverges from the intended architecture. Izurieta and Bieman[16] used "design decay" to express the deterioration of internal structure of system design, while they focused more on the decay of design patterns. Besides, some other terms are also used to describe the same phenomenon, such as design erosion,[17] architecture degeneration,[8] and code decay.[18]

In this SMS, we proposed and refined the definition of architecture erosion: *architecture erosion happens when the implemented architecture violates the intended architecture with flawed internal structure or when architecture becomes resistant to change.* The intended architecture means the conceptual architecture, which is also called *planned architecture*, *as-designed architecture*, and *prescriptive architecture*; the implemented architecture refers to the concrete architecture, which is also named *as-implemented architecture*, *as-built architecture*, *as-realized architecture*, and *descriptive architecture*.[19] Note that the intended architecture is not a static architecture and it can evolve as new requirements emerge; therefore, the intended architecture refers to the currently desired architecture instead of the initially documented architecture. Besides, AEr does not mean temporary violations of design rules in one or two versions but refers to a decreasing tendency of the health status of software systems at the architecture level.

### 2.2 | Characteristics of architecture erosion

AEr occurs in the software development life cycle, which has the following characteristics:

(1) AEr can exist in different phases of the software development process. AEr in software systems is one of the main reasons and manifestations of software erosion and aging.[3,20] The maintenance and evolution phases account for a large part of the life cycle of software development; therefore, a common view is that erosion starts to creep into a system during these two phases. Actually, the seed of erosion might be sowed in the system once architecture patterns were chosen and AEr may already exist before implementing the detailed design (e.g., references[21,22]).

(2) The existence of AEr does not depend on the size of a software system. Some practitioners thought that AEr could only occur in large software systems with complex structures and dependencies, which are harder to understand, but AEr can also exist in small systems.[23]

(3) AEr has an "incubation period." AEr may have an "incubation period" and be a long-term process;[24] that is, when the architecture of a software system suffer from erosion intentionally (e.g., incurring various technical debt for short-term benefits) or unintentionally (e.g., unconsciously breaking the design principles or architectural constraints), the effects of the potential violations may not emerge immediately, such as drastically decreasing the system performance. Sometimes, a well-performed software system could already have an eroded internal structure, and the performance is not the only indicator to judge whether AEr happened or not.[3]

(4) Uncertainty in the speed of AEr. Due to the possible "incubation period," there is a conjecture that AEr is a progressive process. Meanwhile, AEr can also happen quickly in software systems.[23] For example, during the software development process, if newly added components or modifications violate the communication principles among modules specified in design documents, the maintainability of the software system can slide rapidly.

## 3 | MAPPING STUDY DESIGN

This SMS was designed according to the guideline for systematic mapping studies proposed by Petersen et al.[12] The design of the SMS is described in the five following subsections: (1) research questions, (2) pilot search and selection, (3) formal search and selection, (4) data extraction, and (5) data synthesis.

### 3.1 | Research questions

The goal of this SMS formulated based on the Goal–Question–Metric (GQM) approach[25] is to **analyze** the primary studies **for the purpose of** analysis and categorization **with respect to** the concept, symptoms, reasons, consequences, detecting, handling, and lessons learned of AEr **from the point of view of** researchers **in the context of** software development. The goal is further decomposed into eight Research Questions (RQs) (see Table 1) in order to get a comprehensive view of AEr, and the answers of these RQs can be directly linked to the goal of this SMS.

### 3.2 | Pilot search and selection

Before the formal search and selection, we conducted a pilot search and selection to address the potential considerations in the formal SMS (i.e., search terms and time period). These two considerations are elaborated in the following paragraphs.

First, the pilot search can help us settle the appropriate search terms for the formal search.[12] As mentioned in Section 2.1, we realized that the AEr phenomenon is described in various terms in the literature. Therefore, we selected the synonyms of "erosion" (i.e., topic-related terms) from related studies (see Section 2.1) (e.g., "decay," "deterioration," and "degradation") and formulated a trial search string: (erosion OR decay OR degrade OR degradation OR deteriorate OR deterioration OR degenerate OR degeneration) AND (architecture OR architectural OR structure OR structural). Initially, to retrieve as many related papers as possible, we excluded "software" in the software architecture search terms, but the retrieved results encompassed a large number of irrelevant studies (e.g., biochemistry and civil engineering). Then, we selected determiners (i.e., area-related terms) and added several domain-specific terms, such as "software," "software system," and "software engineering," and formulated another search string: (software OR software system OR software engineering) AND (architecture OR architectural OR structure OR structural) AND (erosion OR decay OR degrade OR degradation OR deteriorate OR deterioration OR degenerate OR degeneration). We found that it was precise enough to use "software" as the domain-specific term. As for the area-related terms, we firstly selected "architecture," "architectural," "structure," "structural," and "system," and then we decided to add "design" as a complementary term after a discussion and a trial search, because the number of retrieved results showed a slight increase, which is not a burden to the selection. More details of the formal search terms are presented in Section 3.3.5.

Second, regarding the time period, we initially wanted to choose the search period starting from 1992 when Perry and Wolf published the seminal paper on software architecture and coined the concept of architecture erosion.[1] We did a trial search in the IEEE Xplore database using the query expression ([architectural decay OR architecture decay OR architectural erosion OR architecture erosion OR architectural degradation OR architecture degradation] AND software). The results (see Figure 1) showed that the total number of papers retrieved was 329, where the number of papers published between 1992 and 2005 was 69 (i.e., 21%) whereas the rest (i.e., published between 2006 and May 2019) was 260 (i.e., 79%). In addition, in 2006, Shaw and Clements[27] published the milestone paper about the golden age of software architecture in

**TABLE 1** Research questions and their rationale

| Research questions | Rationale |
|---|---|
| RQ1: What are the definitions of architecture erosion in software development?<br>• RQ1.1: Which terms are used to describe architecture erosion in the definitions?<br>• RQ1.2: What perspectives do the architecture erosion definitions concern about? | Researchers may have different definitions or understanding about the phenomenon of AEr in software development, which are described by various terms and from different perspectives. This might lead to ambiguous interpretation of AEr phenomenon. RQ1 aims to examine which terms are commonly used to define AEr phenomenon and understand AEr phenomenon from different perspectives. By answering RQ1, we can shed light on the common understanding of AEr phenomenon and reach a common definition of AEr. |
| RQ2: What are the symptoms of architecture erosion in software development? | Various symptoms (e.g., violations of design decisions[4]) of AEr have been explicitly discussed in the literature, which can be regarded as indicators of AEr. The aim of this RQ is to provide a taxonomy of different AEr symptoms according to their manifestations, and such a taxonomy may provide a base for future work on detecting AEr. |
| RQ3: What are the reasons that cause architecture erosion in software development? | The phenomenon of AEr does not happen coincidentally but rather due to specific reasons (e.g., architectural violation[26]). These have been separately mentioned in literature. By answering this RQ, we want to determine which reasons of AEr are mostly mentioned in the literature. This can increase the awareness for both practitioners and researchers on the reasons of AEr and consequently support future work on proactively preventing AEr. |
| RQ4: What are the consequences of architecture erosion in software development? | AEr can have a serious impact on a software system and diverse consequences for different stakeholders (e.g., increasing cost). This RQ aims at identifying the potential consequences of AEr and their impact on software development, which can raise the awareness of related stakeholders on AEr and expose the possible high risks of AEr. |
| RQ5: What approaches and tools have been used to detect architecture erosion in software development? | Effective approaches and tools have been proposed and used to identify the eroded structure of architecture. Approaches and tools can facilitate the identification of AEr and check the health status of software systems. The answers to this RQ can tell us what feasible approaches and tools can be used to detect AEr and, to some extent, inspire researchers to develop new approaches and tools. |
| RQ6: What measures have been used to address and prevent architecture erosion in software development? | By answering this RQ, we would like to investigate what kinds of measures that have been developed, proposed, and employed in addressing and preventing AEr. The measures may help to prolong the lifetime of systems and save substantial maintenance effort. |
| RQ7: What are the difficulties when detecting, addressing, and preventing architecture erosion? | The answers to this RQ can shed light on the difficulties and limitations on handling AEr during software development. Additionally, being aware of the difficulties and challenges of handling AEr can help to avoid the pitfalls of AEr and provide a starting point for future research. |
| RQ8: What are the lessons learned about architecture erosion in software development? | Lessons learned refer to the experience presented in the primary studies on coping with AEr in software development. The answers to this RQ will help researchers and practitioners to obtain such experience and get familiar with the AEr in software development. |

Abbreviations: AEr, architecture erosion; RQ, research question.

research and practice, which can also be partially reflected from the statistic result in Figure 1. We decided to use the Pareto principle (the 80\20 rule)[28] to segment the time period, because it is a popular segmentation approach that is widely used in software engineering (e.g., references[28,29]). Considering the distribution of the trial search results (i.e., following the 80\20 rule) and the milestone paper,[27] we finally considered 2006 as the starting year of the search period to study AEr. The end search period is settled as May 2019 when we started this SMS.

## 3.3 | Formal search and selection

The search strategy is a critical prerequisite to an SMS or SLR, because a well-established search strategy can help researchers to retrieve complete and relevant studies as many as possible. In this section, we firstly describe the process of the formal search and selection (see Section 3.3.1), and then we present the search strategy employed in this SMS in three parts: (1) search scope (see Section 3.3.4), (2) search terms (see Section 3.3.5), and (3) selection criteria (see Section 3.3.2). The process of this SMS is illustrated in Figure 2.

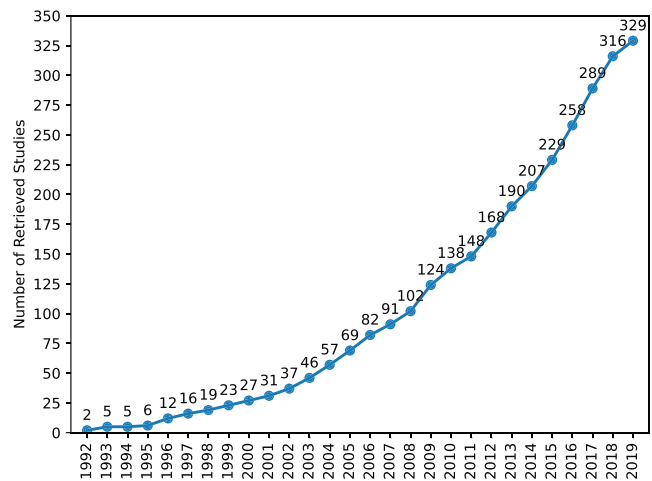

**FIGURE 1** Result of the trial search from 1992 to 2019 using the IEEE Xplore database

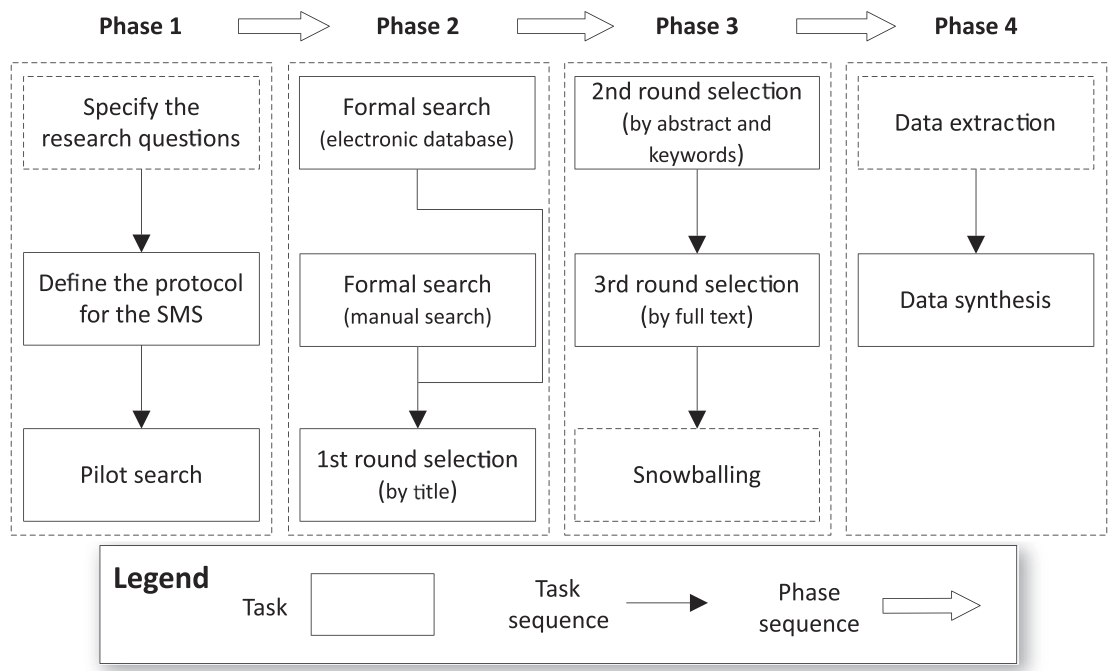

**FIGURE 2** The process of this systematic mapping study

### 3.3.1 | Process

The execution process of this SMS in four phases is shown in Figure 2, and we conducted three rounds of study selection (described in Section 3.3.3).

**Phase 1**: The first author specified the RQs (see Section 3.1) and formulated the protocol for this SMS, which were reviewed by the other authors. Then first two authors conducted a pilot search to determine the search scope (see Section 3.3.4) and search terms (see Section 3.3.5).

**Phase 2:** The first author applied the search scope (see Section 3.3.4) to conduct the study search in the seven electronic databases (see Table 2) and the manual search in the supplementary venues (see Table 3). The first round selection was conducted in this phase and the included studies of the search results were merged by removing the duplicated studies.

**Phase 3**: The first author conducted the second round selection (by abstract) and the third round selection (by full paper). For the results of the third round selection, any uncertain studies were discussed to reach a consensus between the first and second authors about whether the studies should be included or not. After that, the first author conducted a snowballing in this phase, which followed the same three rounds of selection (see Section 3.3.3).

TABLE 2   Electronic databases included in this systematic mapping study

| # | Database | Search scope in database |
|---|---|---|
| DB1 | ACM Digital Library | Title, abstract |
| DB2 | EI Compendex | Title, abstract |
| DB3 | IEEE Explore | Title, keywords, abstract |
| DB4 | ISI Web of Science | Title, keywords, abstract |
| DB5 | Springer Link | Title, abstract |
| DB6 | Science Direct | Title, keywords, abstract |
| DB7 | Wiley InterScience | Title, abstract |

TABLE 3   Journals, conferences, and workshops included in this systematic mapping study

| # | Journals, conferences, and workshops |
|---|---|
| J1 | ACM Transactions on Software Engineering and Methodology (TOSEM) |
| J2 | IEEE Transactions on Software Engineering (TSE) |
| J3 | Empirical Software Engineering (ESE) |
| J4 | IEEE Software |
| J5 | Information and Software Technology (IST) |
| J6 | Journal of Systems and Software (JSS) |
| J7 | Science of Computer Programming (SCP) |
| J8 | Software Quality Journal (SQJ) |
| J9 | Software Practice and Experience (SPE) |
| C1 | International Conference on Software Engineering (ICSE) |
| C2 | International Conference on Automated Software Engineering (ASE) |
| C3 | ACM Joint Meeting on European Software Engineering Conference and Symposium on the Foundations of Software Engineering (ESEC/FSE) |
| C4 | International Symposium on Empirical Software Engineering and Measurement (ESEM) |
| C5 | International Conference on Software Engineering and Knowledge Engineering (SEKE) |
| C6 | European Conference on Software Architecture (ECSA) |
| C7 | Working IEEE/IFIP Conference on Software Architecture (WICSA) |
|  | International Conference on Software Architecture (ICSA) |
| C8 | IEEE International Conference on Software Maintenance (ICSM) |
|  | IEEE International Conference on Software Maintenance and Evolution (ICSME) |
| C9 | International Conference on Quality Software (QSIC) |
|  | International Conference on Software Quality, Reliability and Security (QRS) |
| C10 | IEEE International Conference on Software Analysis, Evolution, and Reengineering (SANER) |
|  | European Conference on Software Maintenance and Reengineering (CSMR) |
|  | IEEE Working Conference on Reverse Engineering (WCRE) |
| W1 | Workshop on Software Architecture Erosion and Architectural Consistency (SAEroCon) |
| W2 | International Workshop on Sharing and Reusing Architectural Knowledge (SHARK) |

**Phase 4**: The first author extracted data from the selected studies (see Section 3.4) according to the data items defined in Table 4. Then, we synthesized the extracted data to answer the RQs defined in Section 3.1, and we also conducted a pilot extraction to reach an agreement on the controversial data items.

### 3.3.2 | Selection criteria

Before the study selection, the following inclusion and exclusion criteria were discussed and defined by the authors after reaching an agreement. The criteria were formulated to select relevant studies for answering the RQs (see Section 3.1) in this SMS. Note that the selection

TABLE 4  Data items extracted from selected studies

| # | Data item name | Description | Relevant RQ |
| --- | --- | --- | --- |
| D1 | ID | The ID of the study. | Overview |
| D2 | Title | The title of the study. | Overview |
| D3 | Author list | The authors' full names of the study. | Overview |
| D4 | Type of authors | The type of authors (i.e., academia, industry, and both). | Overview |
| D5 | Publication type | The type of the study (i.e., journal, conference, workshop, or book chapter). | Overview |
| D6 | Publication venue | The name of the venue where the study was published. | Overview |
| D7 | Publication year | The publication year of the study. | Overview |
| D8 | Definition | The definitions of architecture erosion. | RQ1 |
| D9 | Symptom | The mentioned symptoms can be regarded as indicators of architecture erosion. | RQ2 |
| D10 | Reason | The reasons that lead to architecture erosion. | RQ3 |
| D11 | Consequence | The consequences caused by architecture erosion. | RQ4 |
| D12 | Approach | The approaches used to detect architecture erosion. | RQ5 |
| D13 | Tool | The tools used to detect architecture erosion. | RQ5 |
| D14 | Prevention and remediation | The measures used to address and prevent architecture erosion. | RQ6 |
| D15 | Difficulty | The difficulties when detecting, addressing, and preventing architecture erosion. | RQ7 |
| D16 | Lesson learned | The lessons learned about architecture erosion. | RQ8 |

Abbreviation: RQ, research question.

criteria were applied in all study selection tasks, including pilot search, electronic databases, manual search, and snowballing (see Figure 2).

**(1) Inclusion criteria**

- I1: The paper has been peer-reviewed and is available in full text.
- I2: The paper is related to software development and software architecture.
- I3: The paper mentioned the phenomenon or reasons about AEr, as well as the related measures to handle AEr.

**(2) Exclusion criteria**

- E1: The paper is not written in English.
- E2: The paper is gray literature (e.g., technical reports).
- E3: The content of the paper is only an abstract.
- E4: The paper only mentioned "architecture erosion", but it cannot help to answer the RQs.
- E5: If there are two papers about the same work that were published in different venues (e.g., conference and workshop), the less mature one will be excluded.

### 3.3.3 | Details of the study selection

**The first round selection**: The first author applied the selection criteria (see Section 3.3.2) to identify the primary studies. In the first round, the first author filtered studies based on the titles to select potential primary studies. In order to mitigate potential bias during study selection, we randomly selected 100 papers as a sample from the search results and the first two authors independently conducted a study selection. The inter-rater agreement on the first round selection in the Cohen's Kappa coefficient[30] was 0.82. Any uncertain studies (i.e., they could not be decided by the titles) were temporarily included and kept in the second round selection.

**The second round selection**: The first author selected studies left in the first round selection by reading their abstracts and keywords. The potentially related studies were selected based on the selection criteria (see Section 3.3.2). Similarly to the first round selection, to mitigate potential bias, the first two authors selected independently the remaining 52 papers left in the sample and the Cohen's Kappa coefficient was 0.64. Note that the difference in this round is due to one researcher taking a conservative policy in study selection with the purpose of not missing possibly relevant studies. Any uncertain studies (i.e., those that could not be decided by abstracts and keywords) were temporarily included and kept in the third round selection. All the studies from the sample that were included by one researcher and excluded by the other in this round were eventually excluded in the third round.

**The third round selection**: The first author selected studies by reading the full text of the papers left in the second round selection. The potentially related studies were decided whether they should be finally selected based on the criteria (see Section 3.3.2). Again, to mitigate potential bias in this round, the first two authors independently read the full text of the remaining 26 papers left in the sample and the Cohen's Kappa coefficient was 0.92. Note that, in this final round, any uncertain judgments were discussed together and reached an agreement among all the authors.

**Snowballing**: To collect more potentially relevant studies about AEr, we employed the "snowballing" method[31] to avoid missing any AEr-related studies. Snowballing contains two strategies: backward snowballing and forward snowballing. Backward snowballing refers to the review of the reference list of the included papers, and forward snowballing means to identify the citations to the included papers.[31] In addition, snowballing is an iterative process: the first author checked the references and citations of the studies selected in the third round selection and then re-rechecked the newly included studies by references and citations. This iterative process could stop until there is no any newly selected study. We conducted forward and backward snowballing in Phase 3 (see Figure 2), and each iteration entails the aforementioned three rounds selection (see Section 3.3.3). The newly selected studies in the snowballing process were merged into the final results of the study selection.

### 3.3.4 | Search scope

**(1) Time period**

As illustrated in Section 3.2, we conducted a pilot search to decide the time period of this SMS. After that, we specified the time period of the published studies between January 2006 and May 2019 (i.e., the starting time of this SMS).

**(2) Electronic databases**

The electronic databases selected in this SMS are listed in Table 2; these are regarded as the most common and popular databases in the software engineering field and the selected databases are considered appropriate to search relevant studies.[32] We excluded Google Scholar in this SMS, because the precision of the search results was not acceptable and might include many duplicated and irrelevant results. Additionally, the search results of Google Scholar have overlapped with the seven electronic databases and might omit many relevant papers (e.g., newly published studies).

**(3) Manual search**

To retrieve as many relevant studies as possible, besides the seven electronic databases, we also selected 9 journals, 10 conferences (merged conferences were only counted once, such as WICSA and ICSA), and 2 workshops as supplementary sources to the electronic databases search (see Table 3). These supplementary sources are reputable venues that publish research on software engineering and software architecture and are commonly used in related SMSs and SLRs (e.g., references[33,34]). Note that the selected journals, conferences, and workshops may not be complete, because they were selected as supplementary rather than exhaustive sources.

### 3.3.5 | Search terms

In this SMS, we defined the search terms according to the PICO criteria (i.e., Population, Intervention, Comparison, and Outcome).[13] The population in this SMS is the primary studies on software engineering. The intervention is about the topic of "architecture erosion," which can be divided into two types of search terms (i.e., area-related terms and topic-related terms). Through the pilot search and selection (see Section 3.2), we chose the area-related terms "architecture," "architectural," "system," "structure," "structural," and "design," and the topic-related terms "decay," "erosion," "erode," "degrade," "degradation," "deteriorate," "deterioration," "degenerate," and "degeneration." Finally, the following query string was used in the formal search. We used Boolean OR to join topic-related terms and area-related terms (i.e., synonyms), respectively. We used Boolean AND to join the major terms. The Boolean expression for retrieving relevant studies in database search is as follows:

("software") AND ("architecture" OR "architectural" OR "system" OR "structure" OR "structural" OR "design") AND ("decay" OR "erosion" OR "erode" OR "degrade" OR "degradation" OR "deteriorate" OR "deterioration" OR "degenerate" OR "degeneration").

## 3.4 | Data extraction

To answer the RQs presented in Section 3.1, we extracted and analyzed the data according to the data items (i.e., D8–D16) from each included study. The 16 data items were discussed and formulated by all the authors (see Table 4), which also shows the relevant RQs that are supposed to be answered by the extracted data according to the specific data items.

Before the formal data extraction, we discussed the meaning of each data item and the way of data extraction. To ensure an unambiguous understanding of the data items, the first three authors conducted a pilot data extraction with five studies. The first author extracted data based on the data items from the selected five studies, while the second and third authors reviewed the extracted data. Furthermore, we selected another five studies to conduct a sample data extraction by the first and second authors independently. By comparing the sample data extraction results from the two authors, we established that they largely overlap. We provided both the pilot data extraction results and the sample data extraction results in the replication package.[35] Finally, all divergences and ambiguities of the results of the extracted data were discussed together for reaching an agreement. Likewise, in the formal data extraction process, the data extraction was performed by the first author and reviewed by the second and third authors. Besides, in the process of data synthesis and classification, the extracted data were repeatedly reviewed by the second and third authors and any disagreements were discussed between the first three authors. In this way, we can ensure that the extracted data in the formal data extraction process are valid.

## 3.5 | Data synthesis

In this process, we conducted data synthesis with the extracted data (see Table 4). We employed the descriptive statistics and Constant Comparison method[36] to analyze the qualitative data for answering the eight RQs (see Section 3.1). Note that we employed both descriptive statistics and Constant Comparison as the data analysis methods for answering RQ1, RQ3, RQ4, RQ5, and RQ6. Besides, we also provided examples to clarify the data synthesis results.

Descriptive statistics can provide quantitative summaries based on the initial description of the extracted data; specifically, they were used to analyze the definitions of AEr (i.e., D8), symptoms of AEr (i.e., D9), approaches and tools (i.e., D12 and D13), difficulties (i.e., D15), and lessons learned (i.e., D16). These data can be used to answer RQ1, RQ2, RQ3, RQ4, RQ5, and RQ6, respectively. For instance, we extracted the descriptions of the approaches used to detect AEr from the selected studies to classify the approaches into several categories (see Section 4.6) for answering RQ5. We followed the guidelines of constant comparison[36,37] to analyze the extracted data (see Table 5) for answering RQ1, RQ3, RQ4, RQ5, RQ6, RQ7, and RQ8.

In this SMS, constant comparison, which is a qualitative data analysis method to develop a grounded theory by consistently comparing with the existing findings, was used to generate concepts and categories through a systematic analysis of the qualitative data, including the definitions of AEr (i.e., D8), reasons of AEr (i.e., D10), consequences of AEr (i.e., D11), approaches of AEr detection (i.e., D12), measures of AEr prevention and remediation (i.e., D14), difficulties (i.e., D15), and lessons learned (i.e., D16). The data can be used to answer RQ1, RQ2, RQ3, RQ4, RQ6, RQ7, and RQ8, respectively. The process of constant comparison consists of three steps: (1) Initial coding, executed by the first author, examining the extracted data line by line to identify the topics of the data. For example, "*erosion can happen as a result of many different factors, e.g., there can be a lack of architectural documentation, a lack of developer knowledge, …*" was labeled as "reason" (i.e., D10). (2) Focused coding, executed by the first author and reviewed by the second and third authors, selecting categories from the most frequent codes and using them to categorize the data. For example, "*architectural erosion is known as having a negative impact on the system quality, such as maintainability, evolvability, performance, and reliability, …*" is regarded as a type of "quality degradation" and we mapped the collected quality attributes to the software product quality model (i.e., the ISO/IEC 25010 standard[38]). (3) The disagreements on the coding results were checked and settled down by the first three authors to reduce potential bias and ambiguity. For example, for answering RQ3, we initially coded a reason that causes AEr "*this problem is at

TABLE 5   Relationship between data items, data analysis method, and research questions

| # | Data item name | Data analysis methods | RQs |
|---|---|---|---|
| D1–D7 | Publication information of selected studies | Descriptive statistics | Overview |
| D8 | Definition of AEr | Descriptive statistics and constant comparison | RQ1 |
| D9 | Symptom of AEr | Descriptive statistics | RQ2 |
| D10 | Reason of AEr | Descriptive statistics and constant comparison | RQ3 |
| D11 | Consequence of AEr | Descriptive statistics and constant comparison | RQ4 |
| D12 | Approach | Descriptive statistics and constant comparison | RQ5 |
| D13 | Tool | Descriptive statistics | RQ5 |
| D14 | Measure of prevention and remediation | Descriptive statistics and constant comparison | RQ6 |
| D15 | Difficulty | Constant comparison | RQ7 |
| D16 | Lesson learned | Constant comparison | RQ8 |

Abbreviations: AEr, architecture erosion; RQs, research questions.

*least partially caused by developers' lack of underlying architectural knowledge*" as "*knowledge vaporization*". Then we discussed and agreed that this reason better fits into "understanding issue", and finally, we got 13 categories of reasons as shown in Table 10.

To facilitate the replicability of our SMS, we provide the replication package of this SMS as an online resource.[35] The replication package includes the detailed information of the 73 selected studies (see Appendix A1) and the extracted data based on the 16 data items listed in Table 4.

## 4 | RESULTS

We present the number of searched and selected papers in Section 4.1.1, the demographic data of the selected studies in Section 4.1.2, and the results of the eight RQs from Sections 4.2 to 4.9.

### 4.1 | Overview

#### 4.1.1 | Number of searched and selected papers

The overview of the search, selection, and snowballing results is shown in Figure 3. In the study search phase, we collected 41,723 papers, including 20,551 papers from the 7 databases (see Table 2) and 21,172 papers from the 21 venues (see Table 3). In total, 1220 papers were retained after the first round selection by title, 982 papers were left after removing duplicated papers, 492 papers were kept after the second round selection by abstract, and 68 papers were selected after the third round selection by full text (see Section 3.3.2). Then, we conducted snowballing iterations (see Section 3.3.3) (including forward and backward snowballing) also with the three rounds selection, and five more papers were added by snowballing. Ultimately, 73 (i.e., 68 + 5) papers were finally selected (see Appendix A1).

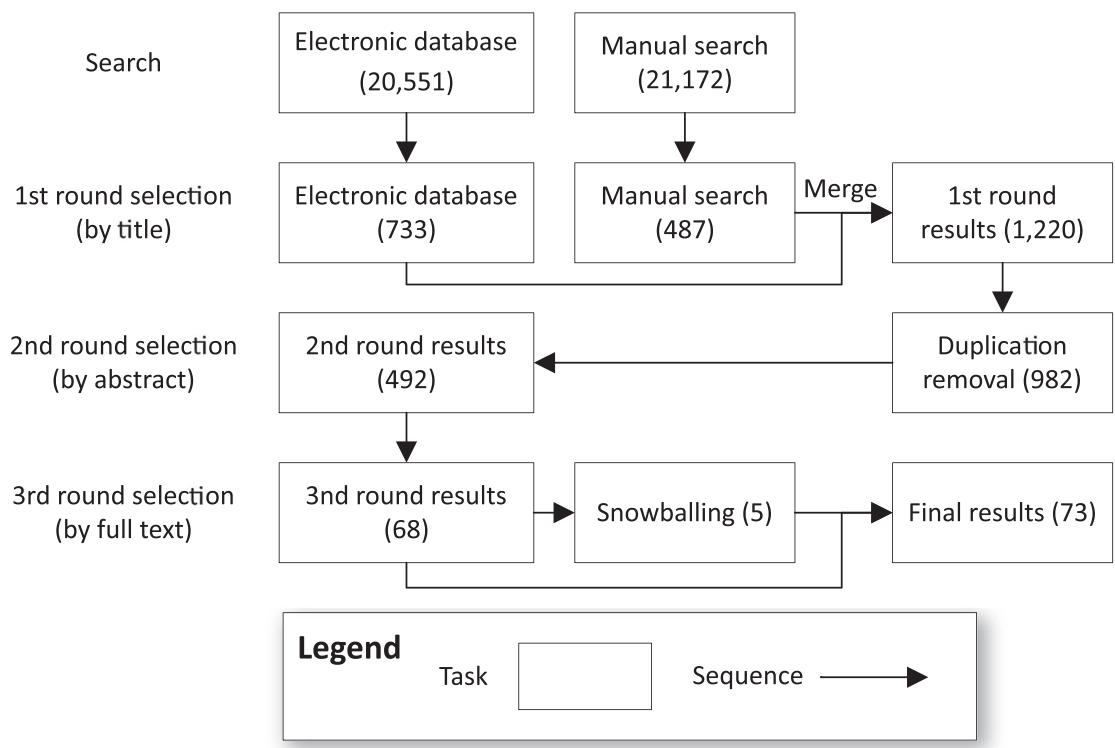

**FIGURE 3** Results of search, selection, and snowballing in this systematic mapping study

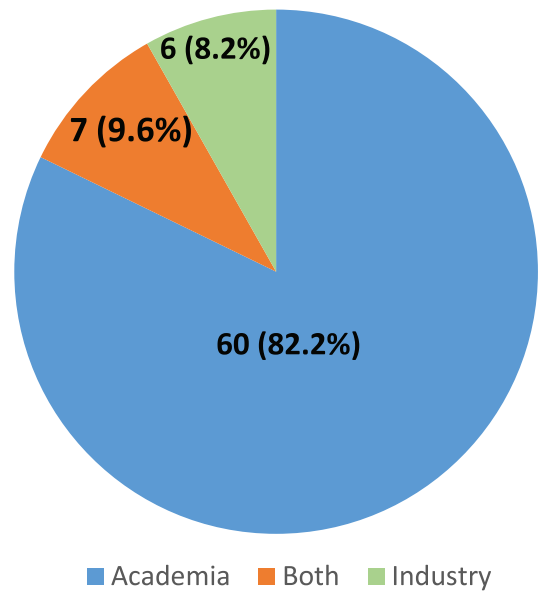

**FIGURE 4** Distribution of the selected studies over types of authors

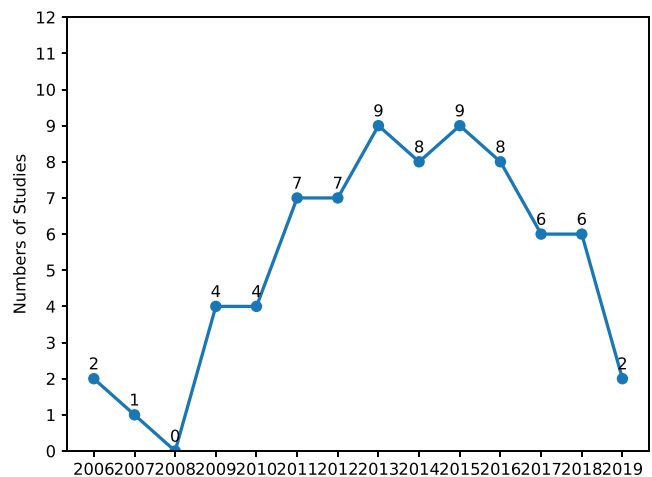

**FIGURE 5** Number of the selected studies over time period

### 4.1.2 | Demographic results

We present the statistical information of the selected studies, including types of authors (see Figure 4), publication years (see Figure 5), and publication venues (see Figure 6).

Figure 4 shows that 82.2% of the selected studies (60 out of 73) are authored by researchers from academia, authors of 9.6% of the selected studies (7 out of 73) work in industry, and the rest (8.2%, 6 out of 73) of the selected studies are collaboration outcomes between academia and industry. Additionally, Figure 5 shows the distribution of the selected studies over the last 13 years (i.e., from 2006 to 2019), from which we can see a fair amount of attention on AEr from 2011 to 2018 (6–10 papers were published per year) with the peak year around 2013. Note that only two papers published in 2019 were included because we stopped our literature search by May 2019.

As shown in Table 6, we list the venues where the selected studies were published, including their names, types, and counts. The 73 studies are published in 49 publication venues (note that merged conferences were only counted once, e.g., WICSA and ICSA were counted as one conference). The leading venues are SANER (six studies including CSMR and WCRE), ICSA (six studies including WICSA and ICSA), ECSA (five studies), ICSE (three studies), JSS (three studies), and IEEE Software (three studies). Figure 6 provides the distribution of the publication venues of the selected studies. Most of the studies were published in conferences (58.9%, 43 out of 73), followed by journals (27.4%, 20 out of 73), workshops (9.6%, 7 out of 73), and book chapters (4.1%, 3 out of 73).

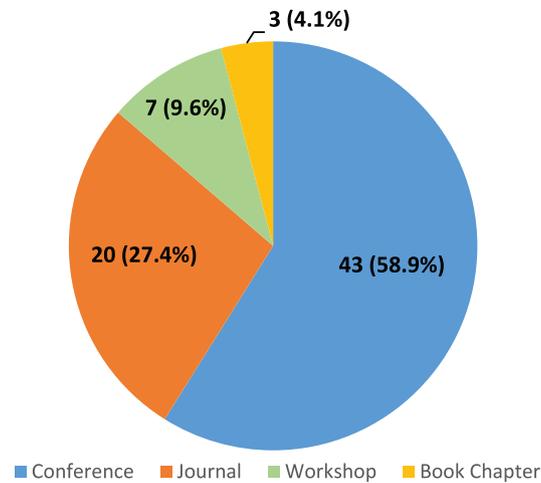

**FIGURE 6** Distribution of the selected studies over types of publication

## 4.2 | RQ1: What are the definitions of architecture erosion in software development?

### 4.2.1 | RQ1.1 Terms of different definitions

As mentioned in Section 2.1, the concept of AEr was proposed by Perry and Wolf[1] in 1992, which is the earliest study that provided the definition of AEr. In the last decades, many studies described the phenomenon of AEr with different terms, such as architectural decay (e.g., [S64]), architecture degeneration (e.g., [S1]), design erosion (e.g., [S72]), and code decay (e.g., [S28]). Table 7 presents all the terms that the selected studies used to describe the phenomenon of AEr. We can see that most studies prefer to use "architecture erosion" to describe this phenomenon, followed by the term "architecture decay". The terms, for example, "architecture degeneration", "design erosion", "code decay", and "software erosion" are also used to represent the phenomenon of AEr. The rest terms are only used in one or two studies to describe the eroded architecture, such as design deterioration (e.g., [S9]) and structure decay (e.g., [S3]).

### 4.2.2 | RQ1.2 Perspectives of different definitions

According to the extraction results of data item D8, we analyzed the definitions of AEr and found that researchers described the AEr phenomenon from different perspectives (i.e., understanding AEr phenomenon from different angles). Based on the analysis of the definitions, we classified the definitions into four perspectives (see Table 8).

**Violation** perspective is the most common description of AEr, which denotes that the implemented architecture of a software system violates the design principles or architecture constraints, and consequently results in system problems and the increasing brittleness of the system.[1] For example, the authors of [S57] stated that "*architectural erosion is the process of introducing changes into a system architecture design that violates the rules of the system's intended architecture*". Violations mainly occur in two phases: (1) Design phase. Developers intentionally or unintentionally change the planned structure of the system architecture due to various reasons (e.g., introducing new design decisions), which gives rise to the violations of original design rules in the design phase of software development (e.g., [S17] and [S21]). (2) Maintenance and evolution phase. Developers accidentally or deliberately make modifications (e.g., for short-term benefits, unfamiliar with the dependencies among components) that violate the design time architectural intents during maintenance and evolution activities (e.g., [S47] and [S57]).

**Structure** perspective highlights the structural problems due to constant erosion of the structure of system systems, where the structure of a software system encompasses its components and their relationships to each other.[39] For example, the authors of [S9] mentioned that "*we define

TABLE 6  Number and proportion of the selected studies over publication venues

| Publication venue | Venue type | Count |
| --- | --- | --- |
| International Conference on Software Analysis, Evolution, and Reengineering (SANER) | Conference | 6 |
| European Conference on Software Maintenance and Reengineering (CSMR) | | |
| Working Conference on Reverse Engineering (WCRE) | | |
| Conference on Software Maintenance, Reengineering, and Reverse Engineering (CSMR-WCRE) | | |
| International Conference on Software Architecture (ICSA) | Conference | 6 |
| Working IEEE/IFIP Conference on Software Architecture (WICSA) | | |
| European Conference on Software Architecture (ECSA) | Conference | 5 |
| International Conference on Software Engineering (ICSE) | Conference | 3 |
| Journal of Systems and Software (JSS) | Journal | 3 |
| IEEE Software | Journal | 3 |
| International Conference on Modularity (MODULARITY) | Conference | 2 |
| Workshop on Software Architecture Erosion and Architectural Consistency (SAEroCon) | Workshop | 2 |
| Software: Practice and Experience (SPE) | Journal | 2 |
| IEEE Transactions on Software Engineering (TSE) | Journal | 1 |
| Information and Software Technology (IST) | Journal | 1 |
| Science of Computer Programming (SCP) | Journal | 1 |
| Automated Software Engineering (ASE) | Journal | 1 |
| Journal of Object Technology (JOT) | Journal | 1 |
| Software Quality Journal (SQJ) | Journal | 1 |
| International Journal of Software Engineering and Knowledge Engineering (IJSEKE) | Journal | 1 |
| International Journal of Computer, Control, Quantum and Information Engineering (IJCCQIE) | Journal | 1 |
| Journal of the Brazilian Computer Society (JBCS) | Journal | 1 |
| Electronic Notes in Theoretical Computer Science | Journal | 1 |
| Journal of King Saud University-Computer and Information Sciences (JKSUCIS) | Journal | 1 |
| Methods and Tools | Journal | 1 |
| Relating System Quality and Software Architecture | Book chapter | 1 |
| Agile Software Architecture | Book chapter | 1 |
| Transactions on Foundations for Mastering Change I | Book chapter | 1 |
| Working IEEE/IFIP Conference on Software Architecture and European Conference on Software Architecture (WICSA/ECSA) | Conference | 1 |
| International Symposium on Foundations of Software Engineering (FSE) | Conference | 1 |
| International Conference on Software Maintenance (ICSM) | Conference | 1 |
| IEEE International Conference on Program Comprehension (ICPC) | Conference | 1 |
| IEEE Working Conference on Mining Software Repositories (MSR) | Conference | 1 |
| Asia-Pacific Software Engineering Conference (APSEC) | Conference | 1 |
| International Conference on Software Engineering and Knowledge Engineering (SEKE) | Conference | 1 |
| International Conference on Software and Systems Reuse (ICSR) | Conference | 1 |
| International Conference on Current Trends in Theory and Practice of Informatics (SOFSEM) | Conference | 1 |
| Annual ACM Symposium on Applied Computing (SAC) | Conference | 1 |
| IEEE Working Conference on Software Visualization (VISSOFT) | Conference | 1 |
| International Doctoral Symposium on Components and Architecture (WCOP) | Conference | 1 |
| International Conference on Advanced Software Engineering and Its Applications (ASEA) | Conference | 1 |
| Annual International Conference on Aspect-oriented Software Development (AOSD) | Conference | 1 |
| International Conference on Software Engineering and Data Engineering (SEDE) | Conference | 1 |

**TABLE 6** (Continued)

| Publication venue | Venue type | Count |
|---|---|---|
| IEEE Prognostics and System Health Management Conference (PHM) | Conference | 1 |
| Annual India Software Engineering Conference (ISEC) | Conference | 1 |
| Conference on Pattern Languages of Programs (PLoP) | Conference | 1 |
| Annual ACM SIGPLAN Conference on Object-Oriented Programming, Systems, Languages, and Applications (SPLASH/OOPSLA) | Conference | 1 |
| IEEE Symposium on Computational Intelligence in Dynamic and Uncertain Environments (CIDUE) | Conference | 1 |
| International Conference on Advances in ICT for Emerging Regions (ICTer) | Conference | 1 |
| International Workshop on Managing Technical Debt (MTD) | Workshop | 1 |
| International Workshop on Modeling in Software Engineering (MiSE) | Workshop | 1 |
| International Workshop on Establishing the Community-Wide Infrastructure for Architecture-Based Software Engineering (ECASE) | Workshop | 1 |
| International Conference on Software Architecture Workshops (ICSAW) | Workshop | 1 |
| Workshop on Sustainable Architecture: Global Collaboration, Requirements, Analysis (SAGRA) | Workshop | 1 |

**TABLE 7** Terms that the selected studies mentioned to describe architecture erosion

| Term | Studies | Count |
|---|---|---|
| Architecture erosion | [S2][S3][S4][S6][S7][S8][S12][S13][S14][S18][S20][S22][S23][S24][S25][S26][S27][S29][S31][S32][S33][S34][S35][S37][S39][S40][S43][S45][S47][S48][S50][S51][S52][S53][S55][S57][S58][S60][S61][S63][S65][S66][S67][S69][S70][S71] | 46 |
| Architecture decay | [S7][S11][S15][S16][S17][S19][S21][S37][S42][S44][S49][S50][S64][S68] | 14 |
| Architecture degradation | [S7][S10][S18][S26][S31][S39][S45][S48][S57][S67] | 10 |
| Architecture degeneration | [S1][S3][S27][S28][S36][S44][S50] | 7 |
| Design erosion | [S1][S9][S15][S18][S55][S62][S72] | 7 |
| Software erosion | [S41][S54][S59][S73] | 4 |
| Code decay | [S28][S56] | 2 |
| Modularity/modular structure deterioration | [S5][S46] | 2 |
| Design pattern rot/decay/grime | [S9][S30] | 2 |
| Architecture deterioration | [S5] | 1 |
| Design deterioration | [S9] | 1 |
| Structure decay | [S3] | 1 |
| Software decay | [S73] | 1 |
| Software aging | [S5] | 1 |
| System rot | [S72] | 1 |
| Design rot | [S7] | 1 |
| Code rot | [S73] | 1 |

*decay as the deterioration of the internal structure of system designs*". From this perspective, the constant erosion of the structure of architecture is derived from the damage to the structure integrity, such as gradually breaking the encapsulation and dependency relationships (e.g., [S9]).

**Quality** perspective refers to the gradually declining product quality[38] of software systems due to architectural changes or increasing architectural smells[40] (e.g., extraneous connector, ambiguous interface, and component overload). A system with an eroded architecture might also have a good runtime performance[3] that is one quality attribute of software systems specified in the ISO/IEC 25010 standard,[38] while the other quality attributes (e.g., reliability and maintainability) may be negatively affected by AEr. Once the environment or requirements changed, it might bring serious consequences to the software system. For example, the authors of [S34] mentioned that "*we regard erosion as the overall deterioration of the engineering quality of a software system*".

**Evolution** perspective denotes the loss of flexibility at the architectural level, which manifests a state of architecture that is almost stagnated and becoming harder to add more changes than before. In this perspective, researchers find that eroding architecture continuously reduces the maintainability and evolvability of the architecture (e.g., [S21] and [S28]). For example, the authors of [S21] stated that "*when the architecture of a software system allows no more changes to it due to changes introduced in the system over time and renders it unmaintainable*".

**TABLE 8** Understanding of architecture erosion phenomenon from four perspectives

| Perspective | Description | Studies | Count |
|---|---|---|---|
| Violation | The implemented architecture (as-built) violates the intended architecture (as-planned) | [S6][S7][S9][S12][S13][S14][S15][S16][S17][S18][S21][S23][S25][S27][S28][S31][S33][S34][S35][S47][S48][S51][S56][S57][S58][S60][S63][S65][S71][S73] | 30 |
| Structure | Constant erosion of the structure of the system architecture | [S3][S5][S7][S9][S22][S24][S26][S27][S73] | 9 |
| Quality | Quality of the software system gradually declines due to architectural changes | [S1][S30][S34][S59][S63][S72] | 6 |
| Evolution | Decreasing maintainability makes it hard to change the eroded architecture | [S21][S28][S50] | 3 |

We found that certain definitions contain multiple perspectives. For example, the authors of [S21] proposed a definition of AEr from three perspectives: violation, quality, and evolution. The definition of AEr provided by [S21] is "*when concrete (as-built) architecture of a software system deviates from its conceptual (as-planned) architecture [violation perspective] where it no longer satisfies the key quality attributes [quality perspective] that led to its construction OR when architecture of a software system allows no more changes to it due to changes introduced in the system over time and renders it un-maintainable [evolution perspective]*".

## 4.3 | RQ2: What are the symptoms of architecture erosion in software development?

AEr shows a tendency to decrease the engineering quality of a system, during which AEr can be detected by miscellaneous signs. Those signs are explicitly mentioned in the literature that can be regarded as *symptoms* and *indicators* of AEr phenomenon, while the symptoms have not been systematically categorized. Therefore, to better understand various symptoms, we collected the AEr symptoms and classified them into four categories as shown in Table 9.

**Structural symptom** includes various structural problems in software systems. Code anomaly agglomeration (e.g., [S17]) and architectural smell (e.g., undesired dependencies [S20] and cyclic dependencies [S38]) are the most frequently mentioned structural symptoms. For example, the authors of [S26] mentioned that "*several studies confirmed co-occurrences of code anomalies are effective indicators of architectural degradation symptoms*". Another example is that the authors of [S64] mentioned "*we use the (architectural) smells as indicators of architecture erosion and consequently to analyze better the sustainability of a system*". Moreover, modular problems (e.g., deteriorated modules and lack of modularity [S36]) can also be regarded as signs of AEr phenomenon. For instance, the authors of [S43] stated that "*we ignore this case and consider lack of modularity as a symptom of architecture erosion*".

**Violation symptom** is a common type of AEr symptoms, which derive from the most common definition of AEr (see Section 4.2) and refers to the violations of prescribed design decisions or constraints (e.g., abstraction and encapsulation). The violations of design decisions or constraints are regarded as a symptom of AEr, which is mentioned in 10 selected studies (e.g., [S31] and [S57]). In [S57], the authors proposed two types of architectural violations: divergence and absence relationship (see Table 9).

**Quality symptom** is comparatively obvious symptom and often manifests in quality attributes of products that can receive attentions of maintainers, such as high defect rate (e.g., [S21]), losing functionality (e.g., [S22]), and gradually decreasing productivity (e.g., [S7] and [S53]). For example, the authors of [S21] mentioned that "*some common symptoms of decay in software architecture include poor code quality, un-localized changes and regressions, high defect rates*".

**Evolution symptom** covers the symptoms related to the evolution process of software systems. For example, rigidity refers to the tendency that architecture becomes difficult to change, because a simple change might cause a ripple effect across all the coupling classes or components[41] (e.g., [S7]). Another example is that the authors of [S1] mentioned that "*increasing (change) difficulty in further evolution and maintenance is a sign of code decay*".

## 4.4 | RQ3: What are the reasons that cause architecture erosion in software development?

In fact, AEr can be caused by various reasons. In Table 10, we list the 13 reasons according to the data collected from the selected studies, as well as their subtypes, description, and related studies. The results show that 64.4% (47 out of 73) of the studies explicitly mention the reasons that cause AEr and we classified these reasons into 13 categories (see Table 10).

**TABLE 9** Categories of symptoms of architecture erosion

| Type | Subtype | Description | Studies |
|---|---|---|---|
| Structural symptom | Code anomaly agglomeration and architectural smell | Certain structural issues (e.g., code anomaly agglomeration and architectural smells) can be regarded as a sign of AEr | [S5][S7][S17][S18][S20][S21][S27][S31][S36][S38][S41][S47][S57][S64] |
| | Co-occurrence of code smells | Certain patterns of co-occurring code smells tend to be stronger indicators of AEr | [S1][S26][S27][S36] |
| | Modularity problem | It refers to the modular issues, such as deteriorated modularization and lack of modularity | [S5][S27][S43] |
| | Class and modular grime | Grime refers to the increase in the number of harmful relationships (e.g., generalizations, associations and dependencies[7]) | [S9][S41] |
| | Others | The study has a generic description about structural symptoms not included in the other subtypes | [S1][S21][S22][S26][S40] |
| Violation symptom | Divergence | A module or relationship that is in the extracted architecture but not in the intended architecture | [S1][S57] |
| | Absence relationship | A module or relationship that is in the intended architecture but not the extracted architecture | [S57] |
| | Others | The study has a generic description about violation symptoms not included in the other subtypes | [S1][S12][S18][S21][S27][S31][S36][S38][S57][S69] |
| Quality symptom | A gradually lower productivity | Decreasing productivity or investing costs cannot receive higher output | [S7][S53] |
| | Functional erosion | Decreasing functionality in software systems | [S22] |
| | High defect rate | The high portion of defective elements compared with all items produced | [S21] |
| Evolution symptom | Rigidity | The tendency for software to be difficult to change, as a change could cause a cascade of subsequent changes in dependent modules[41] | [S1][S7][S21] |
| | Brittleness | The tendency of the software to break in many places every time it is changed, because the changes might cause break in unexpected ways[41] | [S7] |
| | Immobility | The inability to reuse software from other projects or from parts of the same project[41] | [S7] |
| | Increasing correction cost | Gradually increasing cost for correcting architectural defects | [S1] |

Abbreviation: AEr, architecture erosion.

TABLE 10   Reasons that cause architecture erosion

| Reason | Subtype | Description | Studies |
|---|---|---|---|
| Architecture violation | Architectural rule and constraint | Architecture violation happens by violating architectural guidelines (e.g., dependencies between components) | [S34][S46][S54][S58] |
| | Design decision | Developers mistakenly violate design decisions | [S35][S42][S45] |
| | Others | The study has a generic description about architecture violation not included in the other subtypes | [S13][S21][S23][S27][S28][S31][S32][S50][S62][S66][S73] |
| Evolution issue | Inappropriate architecture change | The changes to a system often erode the fundamental characteristics of the original architecture | [S1][S2][S3][S9][S10][S16][S21][S26][S31][S37][S45][S50][S73] |
| | Uncontrolled evolution process | Uncontrolled evolution process might increase the complexity of system and decrease the maintenance | [S56][S58] |
| | Process change | Architecture changes due to the business and development process change | [S7][S22] |
| | Others | The study has a generic description about evolution issue not included in the other subtypes | [S7][S45] |
| Technical debt | Deadline pressure | Due to deadline pressures, architecture changes or compromises have been taken to system architecture | [S9][S12][S22][S46][S47][S48][S53][S62][S66][S69] |
| | Others | The study has a generic description about technical debt not included in the other subtypes | [S4][S55][S73] |
| Knowledge vaporization | Lack of documentation | Incomplete and unavailable documentation and undocumented knowledge | [S34][S36][S37][S45][S47][S50][S53][S54][S57][S66] |
| | Developer turnover | Knowledge lost due to developer turnover | [S34][S69] |
| Requirement issue | Unforeseen and conflicting requirement | Unforeseen, conflicting, and vague requirements may bring uncertainties to software development and cause conflicting design decisions | [S4][S12][S34][S37][S48][S54][S60][S62][S66][S69] |
| | Constantly changing requirement | Constantly changing requirements could give rise to multiple impact (e.g., dramatically increase the project budget and development schedule) | [S45] |
| Understanding issue | Poor understanding of intended architecture | Lack of a full understanding of a system may let the system conflict with the original design during evolution | [S25][S34][S37][S45] |
| | Lack of architecture knowledge | Developers (e.g., novice) who lacks enough architecture knowledge to support the development and evolution | [S39][S44][S45][S53] |
| Suboptimal design decision | Inflexible design | The architecture has low extensibility and is hard to accommodate changes | [S34][S73] |
| | Others | The study has a generic description about design flaw not included in the other subtypes | [S7][S17][S26][S36][S47] |
| Increasing complexity | Design pattern grime | Increasing unrelated artifacts to original design patterns | [S9][S38] |
| | Others | The study has a generic description about increasing complexity not included in the other subtypes | [S34][S51][S54][S56][S64][S66] |
| Organization issue | Poor developer training | Lack of training and education makes it harder to establish uniform programming styles | [S34][S47] |
| | Lack of maintenance | Lack of long-term maintenance (e.g., commitment) to the projects | [S47][S53] |
| | Poor code review process | Bad quality of code review cannot ensure the code quality | [S45] |
| | Increasing workload | Increasing workload might have a negative impact on the development productivity and software quality | [S47] |
| | Non-architectural-centered practice | Focusing on coding efforts instead of the architecture will doom to suffer architecture erosion | [S61] |
| Communication issue | Lack of communication | Lack of communication between stakeholders | [S12][S25][S34][S47] |

(Continues)

TABLE 10 (Continued)

| Reason | Subtype | Description | Studies |
|---|---|---|---|
| | Miscommunication | Miscommunication includes inaccurate information and misunderstanding | [S12][S34] |
| Environment change | Operational environment change | Application environment changed (e.g., standards changes) | [S9][S73] |
| | Obsolete software and hardware component | Third-party libraries update and become incompatible with the current architecture | [S62] |
| Lack of architecture tools | - | Manually fixing problems probably contributes to AEr due to lack of architecture and design tools | [S23][S34][S40] |
| Iterative software development process | - | Modern iterative software development processes (e.g., agile programming methods) may give rise to the occurrence of AEr sooner | [S34] |

Abbreviation: AEr, architecture erosion.

**Architecture violation**. An architecture gradually erodes when architecture violations are introduced into the system by several ways: (1) Rules governing the dependencies between different levels of the system architecture are violated (e.g., [S31]), such as communication rules and conformance rules. (2) Design decisions and architectural guidelines are not strictly followed (e.g., [S34], [S44], and [S46]), such as presentation layer frameworks.

**Evolution issue**. Architecture could hardly keep intact when evolution happens. As the software maintenance and evolution, certain inappropriate changes could break the original architectural integrity and introduce superfluous dependencies. Some seemingly innocuous architecture changes (e.g., bug-fixings and refactoring) could also cause ripple effects to the system architecture (e.g., [S36]). Uncontrollable evolution process (e.g., [S58]) might increase the risk of AEr and give rise to software failure. For example, AEr could be also caused by technological evolution (e.g., [S7]), including operating systems, hardware, and programming languages.[42]

**Technical debt** is a metaphor reflecting technical compromises that can yield short-term benefits but may hurt the long-term goal of a software system.[34] For example, new products are hastily released to the market because of time pressure (e.g., [S66]). Some shortcuts can change architecturally relevant elements (such as classes, components, and modules) and break architectural integrity (such as breaking the encapsulation rules and introducing undesired dependencies), or intentionally incur technical debt (e.g., [S55]). Such suboptimal operations or techniques might increase potential risks to the system quality (such as reliability and extensibility), thereby leading to AEr. More details of technical debt as a reason for AEr are discussed in Section 5.1.3.

**Knowledge vaporization** could cause a series of problems; for example, a poor understanding of project contexts can hinder knowledge transfer among team members, which is often regarded as one of the reasons that cause AEr. Lack of documentation denotes that the previous knowledge (such as design decisions and their rationales) is not available or documented (such as undocumented design decisions and assumptions), which hinders the subsequent architecture enhancements and modifications (e.g., [S57] and [S66]). Besides, developer turnover[43] aggravates AEr by losing knowledge about system requirements and architectural decisions (e.g., [S50]).

**Requirement issue** challenges the architecture sustainability. For example, newly added or constantly changing requirements may conflict with the intended architectural design decisions (e.g., [S4], [S37], and [S45]). Conflicting requirements that are possibly unforeseen in the early stage, and vague requirements may break the consistency and integrity of the intended architecture and negatively influence the maintainability and extensibility of the system architecture (e.g., [S48]).

**Understanding issue**. Poor understanding of the system architecture stems from various factors, such as poor code readability (e.g., bad code quality and lack of code comments), which may act as a trigger for making wrong changes during maintenance and modification. For example, the authors of [S45] pointed out that AEr can derive from bad practices and developer mistakes when developers change the code without understanding the intended architecture that may be violated. Another example is from [S34] that inadequate understanding of the design documentation and principles (e.g., due to poor training) may be a trigger of AEr. Moreover, lack of architecture knowledge can also lead to the understanding problem; for example, developers who are not aware of previous design decisions or have poor knowledge about the intended architecture [S45] could cause damages to the original decisions and architectural structure [S39].

**Suboptimal design decision** is a kind of design defect that often occurs during the design process. Design defects might creep into a system architecture during the design phase, which can be regarded as a hidden danger to system sustainability, as design defects might not be discovered through static analysis.[14] If the target architecture style is initially a suboptimal choice, then trying to fix the design defects later may be impossible.[44] For example, the authors of [S47] mentioned that many maintenance issues originate from inappropriate design. Another example is that the original architecture is not designed to accommodate potential changes (e.g., low extensibility [S34], [S73]).

**Increasing complexity** usually happens when there are increasing couplings and oversized architecture elements (such as class, component, and module) during evolution. Increasing complexity can reduce the understandability of systems and makes it easier to damage the architectural integrity when changes happen (e.g., [S34] and [S56]), because the clean initial architecture cannot be recognized anymore (e.g., [S66]). For example, the authors of [S54] mentioned that the increase in complexity combined with an evident lack of documentation hinders stakeholder to maintain the design aspects of a system and consequently lead to AEr.

**Organization issue**. According to Conway's law,[45] a bad organization can probably generate bad architecture. Various management problems (e.g., lack of maintenance [S47], [S53]) can generate a bad organization, which finally mirrors negative impact in architecture, including high turnover rate, poor training and education of developers, unfair rewarding and punishment metrics, and incompetent code view process. For example, the authors of [S34] mentioned that poor developer training could be a key trigger of AEr. Consequently, AEr can happen due to various organization and management issues.

**Communication issue**. Both miscommunication and lack of communication can incur AEr during the development process. Miscommunication refers to conveying wrong information between stakeholders, while lack of communication can be unconscious. For example, the authors of [S34] stated that communication is vital during system evolution, which is a common cause of AEr including unavailable or misunderstood design decisions.

**Environment change**. The retention of outdated technologies and software (or hardware) components can reduce architecture evolvability, where the software does not follow the changes and AEr will be incurred very soon (e.g., [S73]). Therefore, when the environment changes, the system architecture should also be revised to adapt to the new operational environment; otherwise, environment changes can contribute to the deterioration and erosion of system designs (e.g., [S9]).

**Lack of architecture tools** is a potential cause of AEr for software systems. For example, the authors of [S24] mentioned that AEr might even more severe in systems implemented in dynamic languages, because certain features of dynamic languages (e.g., dynamic invocations and buildings) make developers more likely to break the planned architecture and these languages suffer from lack of architecture tools.

**Iterative software development process** (e.g., agile development methods and extreme programming) can give rise to the occurrence of AEr sooner, because they place less emphasis on upfront architectural design (e.g., [S34]). Recent studies show that the agile process may not make a project agile,[46] and architecture might become the bottleneck of agile projects and rapidly erode if the developers focus on following the agile process rather than increasing architectural agility.

Table 11 shows that the 13 reasons that cause AEr can be further classified into three categories: (1) technical reason, (2) non-technical reason, and (3) both. Technical reason refers to the reasons related to technical application, and non-technical reason denotes the reasons closely related to the human factors in software engineering. Both means that the type consists of both technical and non-technical reasons. The results in Tables 10 and 11 show that technical reasons of AEr (46.6%, 34 out of 73 studies) derive from design, implementation, maintenance, and evolution phases; non-technical reasons of AEr (24.7%, 18 out of 73 studies) stem from organization and management issues; 15.1% (11 out of 73) of the studies mentioned both technical reasons and non-technical reasons. Additionally, the results also indicate that AEr exists in different phases of software development and interacts with many stakeholders.

**TABLE 11** Reasons of architecture erosion in three categories

| Category | Reason | Studies |
| --- | --- | --- |
| Technical reason | Architecture violation | [S13][S21][S23][S27][S28][S31][S32][S34][S35][S42][S45][S46][S50][S54][S58][S62][S66][S73] |
| | Evolution issue | [S1][S2][S3][S7][S9][S10][S16][S21][S22][S26][S31][S37][S45][S50][S56][S58][S73] |
| | Technical debt | [S4][S9][S12][S22][S46][S47][S48][S53][S55][S62][S66][S69][S73] |
| | Suboptimal design decision | [S7][S17][S26][S34][S36][S47][S73] |
| | Increasing complexity | [S9][S34][S38][S51][S54][S56][S64][S66] |
| | Lack of architecture tools | [S23][S34][S40] |
| Non-technical reason | Knowledge vaporization | [S34][S36][S37][S45][S47][S50][S53][S54][S57][S66][S69] |
| | Understanding issue | [S25][S34][S39][S37][S44][S45][S53] |
| | Organization issue | [S34][S45][S47][S53][S61] |
| | Communication issue | [S12][S25][S34][S47] |
| | Environment change | [S9][S62][S73] |
| Both | Requirement issue | [S4][S12][S34][S37][S45][S48][S54][S60][S62][S66][S69] |
| | Iterative software development process | [S34] |

**TABLE 12** Consequences of architecture erosion

| Category | Type | Subtype | Studies |
|---|---|---|---|
| Quality degradation | Maintainability | Modifiability | [S1][S7][S9][S17][S21][S34][S49][S71][S73] |
| | | Modularity | [S9][S12][S43][S46][S49][S51][S52] |
| | | Reusability | [S13][S24][S33][S47][S49][S53] |
| | | Testability | [S9][S30][S47][S51][S52][S73] |
| | | - | [S9][S13][S16][S21][S24][S25][S29][S33][S42][S46][S47][S54][S62][S63][S69] |
| | Evolvability | - | [S12][S24][S29][S46][S47][S53][S63][S69][S71] |
| | Conceptual integrity of the architecture | Understandability | [S21][S42][S47][S51][S54][S73] |
| | Performance efficiency | Capacity | [S22] |
| | | - | [S9][S29][S30][S34][S54][S62] |
| | Portability | Adaptability | [S9][S13][S15][S30][S33][S69] |
| | Extensibility | - | [S51][S52][S54] |
| | Reliability | - | [S29][S49] |
| | Security | - | [S30] |
| | Usability | Operability | [S29] |
| | | - | [S54] |
| | Others | - | [S7][S9][S15][S34][S40][S53] |
| Architectural defect | Brittleness | - | [S11][S32][S48][S73] |
| | Architecture mismatch | - | [S16][S40][S47] |
| | Anti-pattern | - | [S9] |
| | Poor system integrity | - | [S21] |
| | Others | - | [S1][S9][S15][S21][S34][S54][S71][S73] |
| Increased cost | - | - | [S34][S38][S48][S49][S51][S52][S53][S54][S69][S73] |
| Failure of software projects | Shortened system lifetime | - | [S1] |
| | Others | - | [S7][S13][S33][S34][S44][S54][S69][S73] |
| Failure to meet requirements | - | - | [S7][S48][S73] |
| Software aging | - | - | [S9][S34][S48] |
| Technical debt | Architectural debt | - | [S49] |
| | Others | - | [S42][S49] |
| Organization disintegration | - | - | [S7] |

## 4.5 | RQ4: What are the consequences of architecture erosion in software development?

We identified various consequences of AEr mentioned in the selected studies (see Table 12). These consequences can be classified into eight categories.

**Quality degradation** is the most frequently mentioned consequence and AEr can result in various quality attributes (QAs) degradation. In this SMS, we mapped the degraded QAs according to the ISO/IEC 25010 standard.[38] For example, AEr can give rise to the deterioration of the architectural structure, make the system less flexible (e.g., hard to enhance and extend [S7]), decrease modularity of systems (e.g., reduce the components' cohesion [S50]), and make systems harder to maintain without breaking certain dependencies. For instance, the authors of [S34] mentioned that "*even if a system does not become unusable, erosion makes software more susceptible to defects, incurs high maintenance costs,* *degrades performance and, of course, leads to more erosion*". Besides, stakeholders may not even be able to understand the ramification of breaking these dependencies.

**Architectural defect** denotes that an eroded architecture will make the architecture to have more defects (i.e., defect proneness), for example, anti-patterns [S9] and superfluous dependencies [S47], [S73], which in turn is likely to give rise to more erosion. For example, the authors of [S54] mentioned that "*even if a software system does not become completely inoperative, erosion will make the system more predisposed to defects*". Architecture mismatch[47] is also an architecturally relevant defect arose in eroded software systems (e.g., [S47]). The potential problems of architectural mismatch may derive from the assumptions made to the components, connectors, and global architecture structure.[47]

**Increased cost** denotes that rising costs (including time and labor cost) need to be invested into activities like maintenance and refactoring. Moreover, it may increase the time-to-market of software products and bring about the budget of development cost overrun (e.g., [S51] and [S69]).

**Failure of software projects** might be the worst consequence of AEr. Once uncontrollable AEr has led to irreversible effects and irreparable software systems (or become less cost-effective to be maintained), it marks the failure of software projects and the systems need to go through a process of reengineering and redevelopment (e.g., [S34] and [S44]). Besides, AEr shorten the lifetime of software projects and cause rapid obsolescence, or trigger bottom-up system reengineering (e.g., [S1]).

**Failure to meet requirements** means that AEr can result in the decreasing ability of a system to satisfy the requirements of stakeholders. For example, the authors of [S48] mentioned that the potential consequences of AEr encompass failure to meet the requirements, as changes become difficult to be made on software systems due to various reasons, such as increasing complexity in eroded architecture (e.g., [S7]).

**Software aging** is a phenomenon caused by AEr that refers to the tendency of performance degradation and failure of software systems, and a general characteristic of this phenomenon is that the systems failure rate will increase and many aging-related errors occur (e.g., erroneous outcomes).[20,48] For example, the authors of [S9], [S34], and [S48] mentioned that the low adaptability and extensibility arose in an eroded architecture could further deteriorate the systems, and AEr could accelerate software aging.

**Technical debt** would be incurred in software systems that AEr exists and degrades the quality of software systems, where short-term compromises lead to significantly potential hidden danger to software systems regarding fixing bugs or adding new features (e.g., [S49]). For example, the authors of [S42] mentioned that AEr has been shown to incur technical debt and decrease a system's maintainability; therefore, tracking AEr is critical. Besides, the authors of [S49] stated that "*such an (eroded) system incurs a technical debt, where short-term compromises lead to significant long-term problems in terms of reduced ability of fixing bugs or adding new features*".

**Organization disintegration** denotes that AEr could cause inconsistent management issues, thereby disintegrating the organization. For example, the authors of [S7] argued that the newly hired developers in a development team cannot completely understand the eroded architecture and the workload finally falls on the original developers who might have to work hard and not resist the stress, which may incur high turnover in the long run.

## 4.6 | RQ5: What approaches and tools have been used to detect architecture erosion in software development?

### 4.6.1 | Approaches

As shown in Table 13, we classified the collected approaches of AEr detection into four categories, which are detailed below.

**Consistency-based approach** refers to approaches based on the evaluation of architecture consistency,[49] which aims to align the implemented system with the intended architecture. Consistency-based approaches are one type of the most commonly applied approaches for detecting AEr during the development process; 34.2% (25 out of 73) of the studies propose consistency-based approaches to detect AEr through evaluating architecture consistency, in which around 56.0% (14 out of 25) of the studies use the Architecture Conformance Checking (ACC) techniques to detect AEr through checking the conformance between the implemented architecture and the intended architecture (e.g., [S28], [S38], and [S43]). For example, specific architectural rules can be defined and used to compare the implemented architecture against the specified constraints, such as rules of constraints (e.g., [S33], [S45], and [S53]), formal representation (e.g., [S13] and [S68]), Architecture Description Languages (ADL) (e.g., [S29] and [S48]), and Domain-Specific Language (DSL) (e.g., [S14], [S31], and [S58]). In addition, other approaches can also be used to evaluate architecture consistency. Reflexion Modeling (RM) technique[50] enables developers to extract high-level models and map the modules to source code elements for building high-level models that capture the intended architecture of the system, thereby identifying AEr. Design Structure Matrix (DSM) can establish a dependency matrix among classes of a system without relying on the higher level components, which enables developers to visually check violations for detecting AEr (e.g., [S49] and [S51]).

**Evolution-based approach** is commonly used to identify AEr by checking different history versions of projects. For example, the authors of [S15] proposed the ADvISE approach to analyze the evolution of the architecture at different levels. The authors proposed a representation approach of architecture based on classes and their relationships, and they analyzed the architectural history at various levels (e.g., classes, triplets, and micro-architectures) to identify and measure AEr. Besides, to study architectural changes and erosion, the authors of [S16] conducted an

**TABLE 13** Approaches used to detect architecture erosion

| Category | Approach | Description | Studies |
| --- | --- | --- | --- |
| Consistency-based approach | Architecture Conformance Checking (ACC) | Checking whether the implemented architecture is consistent with the intended architecture (e.g., checking architectural violations) | [S13][S14][S28][S29][S31][S33][S38][S43][S45][S48][S53][S58][S68][S73] |
| | Reflexion Modeling (RM) | It helps to extract high-level models, map the implementation entities into these models, and compare the two artifacts | [S27][S40][S55][S57][S65][S69] |
| | Design Structure Matrix (DSM) | Detecting relations of modules by a creating dependency matrix to visually check for violations | [S49][S51] |
| | Light-weight Sanity Check for Implemented Architectures (LiSCIA) | LiSCIA can help to derive architectural constraints from the implemented architecture and spot potential problems leading to AEr | [S8][S28] |
| | Detection of design pattern grime and rot | Comparing the specified design patterns with realized patterns | [S9] |
| | Detection of dependencies between architectural components | Analyzing the number and types of dependencies between architectural components | [S11] |
| Evolution-based approach | Architectural Decay In Software Evolution (ADvISE) | ADvISE aims to analyze architecture evolution at various levels that helps to analyze how and where AEr occurred | [S15][S56][S61] |
| | Architecture Recovery, Change, and Decay Evaluator (ARCADE) | A software workbench used to conduct architecture recovery, architectural changes metrics, and AEr metrics | [S16][S17] |
| | MORPHOSIS | Detecting AEr through analyzing evolution scenarios, checking architecture enforcement, and tracking architecture-level code metrics | [S51][S52] |
| | Profilo | Detecting AEr by analyzing the history and trends of system evolution | [S56] |
| | Architecture Analysis and Monitoring InfraStructure (ARAMIS) | Detecting AEr by runtime monitoring, interaction validation among architectural components, mapping the captured behavior of architectural elements | [S67] |
| | Variant Analysis (VA) | Detecting and comparing architectural violations across different versions | [S44] |
| Defect-based detection | Defect-based measurement | Investigating the defect–fix history and detecting hotspots that contribute to AEr by analyzing certain defects (e.g., interface defects) | [S1][S73] |
| | Detection of architectural smells | Detecting AEr by detecting architectural smells | [S36][S64] |
| | Software Prognostics and Health Management (PHM) approach | Building a PHM model to identify and predict the health status of software systems based on discriminant coordinates analysis (DCA) | [S3] |
| | Detection of design pattern grime and rot | Comparing the specified design patterns with realized patterns to identify AEr | [S9] |
| Decision-based approach | Capturing design decisions about the adopted architectural patterns | Detecting AEr by capturing the most important design decisions about the adopted architectural patterns | [S37] |
| | Detect and trace architectural tactics | Detecting AEr through identifying and tracing the presence of architectural tactics and keeping developers informed of underlying architecture decisions | [S39] |

Abbreviation: AEr, architecture erosion.

empirical study that they employed the ARCADE approach to analyze the evolution history of the architecture of 14 open-sourced Apache projects using architectural recovery, architectural change metrics, dependency analysis, and evolution paths.

**Defect-based detection** is extensively used to detect AEr, which refers to the review and inspection process of detecting AEr through detecting system defects. For example, the authors of [S1] conducted a case study of defect-based measurement and they found that hotspot components contain 50% more defects that contribute most to AEr. Other approaches can also help to detect AEr by identifying the symptoms of AEr (see Section 4.3). For example, the authors of [S64] proposed the symptoms of AEr (e.g., interface-based smells and dependency-based smells) and AEr metrics to analyze the AEr trend in software systems.

**Decision-based approach** aims to detect AEr by capturing important design decisions, including the selected architectural patterns and tactics. For example, the authors of [S37] detected AEr by redefining a collection of architectural patterns. They leveraged the proposed ABC tool to capture design decisions and generate OCL code to specify the constraints of the patterns, in order to detect AEr during the design phase and validate the runtime architecture.

### 4.6.2 | Tools

Versatile tools can facilitate the understanding of the system architecture and make it more convenient for developers to visualize, identify, track the symptoms of AEr, and finally repair or mitigate AEr. There are 41.1% (30 out of 73) of selected studies that proposed or investigated relevant tools for detecting AEr. In total, 35 tools were collected from the selected studies, and the numbers of open-source tools (51.4%, 18 out of 35) and commercial tools (48.6%, 17 out of 35) are close to each other. Three tools are not available for download now.

The tools have various features and the classification of tools is presented in a similar manner developed for AEr detection approaches according to their main purposes mentioned in the studies. In Table 14, we classified the 35 tools into four categories based on their distinctive purposes. The results in the "URL" column were mainly collected from the selected studies, part of which were collected from the Internet. "Not available" means that we cannot find an available URL link regarding whether the tools are still available now.

**Conformance checking** denotes that this type of tools is used to check architecture conformance for ensuring the developers and maintainers have followed the architectural edicts set and not eroding the architecture by breaking down abstractions, bridging layers, compromising information hiding, and so on.[39] Tools in this type support detection of architectural violations, analysis of dependencies, and so forth. For example, Axivion allows developers and architects to conduct dependency analysis, quality metrics, and visualization for checking architecture conformance by identifying and stopping AEr (e.g., [S18] and [S34]).

**Quality measurement** means that tools of this type provide various metrics to evaluate the quality of systems, monitor software systems, trace decisions, and so forth. For example, Archie is an Eclipse plug-in that helps developers know about the impact of modifications and refactorings on sensitive areas of code, thereby contributing to mitigating the long-term problems related to AEr in the system.

**History analysis** denotes that this type of tools can be used to analyze the evolution trend, trace evolutionary scenarios, and parse the software architecture. For example, Sotograph consists of a suite of tools and supports the analysis of detailed structure, quality, and dependency on different abstraction levels, such as cyclical dependencies and duplicated code blocks (e.g., [S63], [S65], and [S66]).

**Visualization** refers to the tools that employ graphical representation of architectural elements to detect AEr, such as the visualization of anti-patterns and architectural smells (e.g., cyclic dependencies). Visualization techniques contribute to the understanding of the structure of large software systems by graphic or multiple-dimensional representations. For instance, the authors of [S72] demonstrated that Getaviz is an easy-to-use tool for visualizing software evolution, which enables developers to visualize the erosion of system design and architecture, such as the evolution of cyclic dependencies and anti-patterns.

### 4.7 | RQ6: What measures have been used to address and prevent architecture erosion in software development?

We find that the measures used to address and prevent AEr encompass preventive and remedial measures. As shown in Table 15, we classified the measures used to address AEr into three categories: (1) preventive measure, (2) remedial measure, and (3) both. Preventive measures are usually employed in different architecture activities (e.g., architecture analysis and architecture evaluation), and these measures refer to the proactive countermeasures for avoiding AEr or mitigating the risk of AEr. Remedial measures aim at repairing (or improving) the existing eroded parts of the implementation. Both means the measures help to address AEr including the above-mentioned two measures.

### 4.7.1 | Preventive measures

**Architecture Conformance Checking** (ACC) are employed to review the process-based activities during software development for ensuring the implemented architecture aligns with the intended architecture. The authors of [S23] mentioned that "*checking architecture conformance bridges the gap between high-level models of architectural design and implemented program code, and to prevent decreased maintainability, caused by architectural erosion*", which indicates that ACC can help identify and correct architectural violations and further avoid constant AEr during the software

TABLE 14  Tools used to detect architecture erosion

| Type | Tool | URL link | Description | Open source | Studies |
|---|---|---|---|---|---|
| Conformance checking | Lattix | https://www.lattix.com/products-architecture-issues/ | It can be used to extract the architecture dependencies for identifying architectural violations. | No | [S18][S23][S25][S28][S34][S51][S52][S63][S65] |
| | Structure101 | https://structure101.com/ | It offers visualized views of code organization and helps practitioners to better understand the structure and remove cyclic dependencies. | No | [S13][S18][S23][S25][S34][S63][S65][S66] |
| | Axivion | https://www.axivion.com/en | It supports various dependency analyses and helps stop software erosion, analyze, and recover the methods developed for legacy software by understanding the software architecture. | No | [S18][S34][S51][S52][S63][S65][S66] |
| | Sonargraph | https://www.hello2morrow.com/products/sonargraph | It is a static code analyzer that allows developers to monitor a software system for technical quality, enforce architectural rules (e.g., relationships between layers), and detect violated dependencies. | No | [S18][S23][S25][S34][S63] |
| | ArCh | https://wiki.archlinux.org/index.php/Java_package_guidelines | It can be used to check dependencies violations and generate reports about the conformance checking of system architecture. | Yes | [S13][S60][S69] |
| | ConQAT | https://www.cqse.eu/en/news/blog/conqat-end-of-life/ | It can depict all modules as UML components and check the consistency of the defined architectural model. | No | [S23][S63][S68] |
| | SAVE | Not available | It provides a graphical editor to define the intended architecture and show violations after the evaluation. | No | [S23][S44][S65] |
| | ARCADE | https://softarch.usc.edu/wiki/doku.php?id=arcade:start | It supports architecture recovery, architectural change and decay metrics, and helps perform different statistical analyses. | Yes | [S11][S16] |
| | Dclcheck | https://github.com/rterrabh/pi-dclcheck | It can be used to verify whether the implemented architecture respects the specific constraints. | Yes | [S4][S69] |
| | Macker | https://sourceforge.net/projects/macker/ | It can be used to define a diversity of conformance rules and provide violation reports. | Yes | [S23][S63] |
| | ArchRuby | https://aserg.labsoft.dcc.ufmg.br/archruby/manual.html | It can generate reports about the illustration of the architectural conformance and visualization processes. | No | [S24] |
| | DCL2Check | https://github.com/aserg-ufmg/dcl2check | As a plug-in for Eclipse IDE, it supports architectural conformance verification and high-level architectural visualization. | Yes | [S35] |
| | Card | https://code.google.com/archive/p/card-plug-in/downloads | As a plug-in for Eclipse IDE, it can be used to search for UML and Java files, as well as conformance checking. | Yes | [S48] |
| | ReflexML | https://github.com/dPacc/ReflexML | It can be used to create a UML-embedded mapping of architecture models to code and check the consistency (e.g., architecture-to-code traceability information) based on the mapping results. | Yes | [S63] |
| | dTangler | https://github.com/vladdu/dtangler | It helps conduct conformance checking based on the DSM technique that is complemented with text-based editors to define rules. | Yes | [S23] |

TABLE 14 (Continued)

| Type | Tool | URL link | Description | Open source | Studies |
|---|---|---|---|---|---|
| | Coverity | https://www.synopsys.com/software-integrity/security-testing/static-analysis-sast.html | It provides a set of quality metrics and various forms of dependency analysis to check architecture conformance. | No | [S34] |
| | STAN | https://stan4j.com/ | It helps to detect design flaws, analyze dependencies, and visually understand system structure. | No | [S18] |
| | Dependometer | https://sourceforge.net/projects/dependometer/ | It helps validate dependencies against the logical architecture structuring the system into classes, packages, subsystems, vertical slices, and layers and detects cycles between these structural elements. | Yes | [S63] |
| | Classycle | https://classycle.sourceforge.net/ | It allows to define and analyze the dependencies of classes and packages (especially helpful for finding cyclic dependencies). | Yes | [S63] |
| | FIS system | https://www.fispro.org/en/ | It is a fuzzy rule-based system (a.k.a fuzzy inference system) used to detect AEr symptoms and recommend treatments to reduce or mitigate AEr symptoms. | Yes | [S41] |
| | Klocwork | https://www.perforce.com/products/klocwork | It is a static analysis tool that provides support to architecture visualization in the form of graphs, which helps detect security issues and increase code reliability for ensuring architecture conformance. | No | [S34][S65] |
| | ABC | Not available | It provides support to design architecture, adopting predefined architectural patterns, capturing design decisions, and detecting AEr. | Yes | [S37] |
| | SIC | Not available | It helps to extract the descriptive architecture using JavaDoc comments and detect the architecture violations of a codebase change. | Yes | [S45] |
| Quality measurement | JDepend | https://github.com/clarkware/jdepend | It allows to automatically measure the quality of a design in terms of its extensibility, reusability, and maintainability to effectively manage and control package dependencies. | Yes | [S18][S34][S65] |
| | Archie | https://github.com/ArchieProject/Archie-Smart-IDE | As a plug-in for Eclipse IDE, it helps to automate the creation and maintenance of architecturally relevant trace links between code, architectural decisions, and related requirements. | Yes | [S19][S39] |
| | SourceAudit | https://www.frontendart.com/en/ | It can measure the maintainability of source code and automatically monitor the source code quality of software system. | No | [S59] |
| | JArchitect | https://www.jarchitect.com/ | It can provide various metrics to analyze the architecture and generate evaluation reports. | No | [S53] |
| | SonarQube | https://www.sonarqube.org/ | It supports inspecting code quality by performing automatic reviews. | No | [S18] |
| | Xradar | https://xradar.sourceforge.net/ | It is an open extensible code report tool and can generate HTML/SVG reports of the systems' current state and development over time. | No | [S63] |
| History analysis | Sotograph | https://www.hello2morrow.com/products/sotograph | It consists of a number of tools and supports the trend monitoring, quality and dependency analysis on different abstraction levels, as well as the changes in architecture violations between various versions. | No | [S63][S65][S66] |
| | CppDepend | https://www.cppdepend.com/ | (For C++) It is used to measure some basic metrics and trace evolution scenarios to the code and reveal potential ripple effects. | No | [S51][S52] |

(Continues)

TABLE 14 (Continued)

| Type | Tool | URL link | Description | Open source | Studies |
|---|---|---|---|---|---|
| | NDepend | https://www.ndepend.com/ | (For C#) It is used to measure some basic metrics and trace evolution scenarios to the code and reveal potential ripple effects. | No | [S51][S52] |
| | EVA | https://github.com/namdy0429/EVA | EVA helps explore, visualize, and understand multiple facets of architectural evolution. | Yes | [S42] |
| Visualization | Understand | https://scitools.com/ | It helps to understand and maintain the poorly documented legacy systems by visualizing and metrics. | No | [S18] |
| | Getaviz | https://github.com/softvis-research/Getaviz | It is a toolkit that automatically generates the visualization of the eroded part of systems, thereby assisting developers and architects in evaluating, tracing, and combating AEr. | Yes | [S72] |

Abbreviation: AEr, architecture erosion.

TABLE 15 Measures used to address architecture erosion

| Category | Type | Description | Studies |
|---|---|---|---|
| Preventive measure | Architecture conformance checking | Checking and correcting the inconsistencies between the intended architecture and the implemented architecture | [S2][S4][S6][S13][S14][S20][S23][S24][S25][S29][S31][S34][S35][S40][S48][S52][S58][S61][S63][S65] |
| | Architecture monitoring and evaluation | Monitoring and evaluating architectural status (e.g., assessing QAs metrics and dependency analysis) | [S8][S17][S20][S34][S43][S51][S62][S67][S70][S73] |
| | Establishing traceable mechanism | Tracing and managing artifacts (e.g., decisions and tactics) by certain feasible methods and tools (e.g., documentation and links) | [S10][S14][S34][S61] |
| | Evolving architecture as changes occur | Evolving architecture (and architecture model) when changes occurred (e.g., environment change, requirements change, and workflows change) | [S25][S32][S34] |
| | Explicitly defining architecture | Explicitly defining the intended architecture and exposing related knowledge | [S19][S29][S31] |
| Remedial measure | Architecture maintenance | Conducting maintenance activities to repair architectural defects and adapt the system architecture | [S7][S9][S12][S26][S29][S34][S41][S53][S57][S60][S73] |
| | Architecture restoration | Restoring architecture (e.g., reverse engineering and model discovery) | [S50][S53][S54][S71] |
| Both | Architecture refactoring | Refactoring the codebase (e.g., remodularizing and restructuring) | [S5][S18][S45][S46][S62][S69] |
| | Visualization | Architecture visualization can help to understand the architectural structure | [S35][S72][S73] |
| | Management optimization | Optimizing management skills and improving management commitment | [S73] |

Abbreviation: QAs, quality attributes.

development life cycle. In addition, establishing architecture constraints can be used to support ACC. Architecture constraints refer to the rules for the specification and enforcement of architecture, which help to keep the architecture consistent with the specified restrictions, such as constraints for consistency checking between architecture decisions and component models. Establishing architecture constraints contributes to the reliability and robustness of software systems and reduces AEr risks. For example, the authors of [S4] proposed a Dependency Constraint Language (DCL) to restrict the spectrum of architectural dependencies. Additionally, the authors of [S35] proposed a domain-specific language (i.e., DCL 2.0) with a supporting tool (i.e., DCL2Check) to facilitate modular and hierarchical architectural specifications, which helps the development teams to handle architectural violations for addressing AEr.

**Architecture monitoring and evaluation** aims at employing means to monitor the health status of software systems and evaluate whether the symptoms of AEr,[40] such as architectural smells (e.g., extraneous connector, ambiguous interface, and component overload), crept into a system. For example, the authors of [S8] applied LiSCIA, an architecture evaluation method, at different stages of software development and periodically evaluated the implemented architecture, which can produce evaluation reports and guidelines to improve the implemented architecture, thereby helping to identify and prevent AEr. Besides, dependency analysis is a feasible measure for identifying and determining the interdependence between various entities (e.g., classes, packages, modules, and components), which enables developers to understand the relationships and directions of the dependencies. For example, the authors of [S20] proposed a constraints concerning plug-in based on a domain-specific language (i.e., DepCoL) to define constraints regarding plug-ins and feature dependencies. In this way, AEr can be prevented in plug-in-based software systems.

**Establishing traceable mechanism** denotes that available mechanisms should be established to manage and trace software artifacts and support maintenance activities during the software development and evolution process, such as design decisions and tactics. For example, in order to mitigate and avoid pervasive problems of AEr, the authors of [S10] established six trace creation patterns for creating traceability between concrete architectural elements and design decisions, which can keep developers informed about implemented architectural tactics, styles, and design patterns.

**Evolving architecture as changes occur** refers to the measures with the goal of evolving an architecture when changes to the system happen, such as environment, requirements, and workflows changes. For instance, the authors of [S25] claimed that architecture should be built for satisfying all current requirements and incrementally evolved once requirements changed. Although it might be time-consuming, to some extent, these measures are feasible to prevent and minimize AEr. Another example is from [S34], the authors mentioned that "*the contribution of self-adaptation strategies towards preventing architecture erosion relies on minimising human interference in routine maintenance activities.*" Self-adaptation techniques are employed to build and manage the systems by following the principles of autonomic computing, in order to respond to possible changes (e.g., requirement and environment changes) and address uncertainty at runtime.[51] With this measure, architectural problems might be detected and resolved early during the development process, which reduces the demands for making fixes afterwards that are potentially error-prone.

**Explicitly specifying architecture** could partially help to prevent AEr by defining architecture explicitly and exposing underlying design decisions, architecture tactics, and constraints, in order to ensure stakeholders and maintainers completely understand the details of what they will implement and reduce the risk from the implementation phase (e.g., [S19]). For example, the authors of [S29] argued that the AEr control techniques should provide mechanisms for explicitly defining the intended architecture and checking whether the implemented system conforms to the intended design.

### 4.7.2 | Remedial measures

**Architecture maintenance** is a common measure used to repair eroded architecture with the purpose of maintaining architectural sustainability. Architecture repair aims at using repairing strategies to fix up architectural anomalies, such as synchronizing the implementation with intended architecture and repairing detected violations. For example, the authors of [S18] employed tools (e.g., Sonargraph and Structure101) to repair certain architecture problems (e.g., design violations and cyclic dependencies). Another example is that the authors of [S12] proposed an architectural repair recommendation system that provides refactoring guidelines for developers and maintainers to fix architectural violations. Note that inappropriate architectural modifications (e.g., violating architectural rules) could break the system structure over time, while suitable modifications can help to reverse or minimize AEr for the purpose of architecture optimization. For example, the authors of [S9] optimized the architecture by removing design pattern grime and rot, which can potentially reduce maintenance costs and improve system adaptability, consequently stop and remedy the eroding architecture.

**Architecture restoration** typically includes reverse engineering techniques to discover and recover architectural structures from software artifacts, as well as speculate the intended architectural design. For example, the authors of [S50] proposed an approach to regenerate architecture and extract architecturally significant concerns that led to eroded model artifacts, in order to counteract the erosion process.

### 4.7.3 | Both

**Architecture refactoring** aims to improve the structure of a system without changing the external behaviors of the system. For example, the authors of [S46] proposed a multi-objective optimization method for improving the original package structure, preserving the original design decisions, and remedying eroded parts in an architecture. Besides, simplifying architecture is also one of the purposes of architecture refactoring, which aims at deliberately controlling architectural complexity and further simplifying the system during the maintenance and evolution phases. When architectural complexity proliferates towards being uncontrollable, simplifying the system architecture can be useful for understanding the architecture of the system (e.g., [S18] and [S69]), reducing the risk of AEr occurrence from the increasing complexity, and decreasing the difficulty of conformance checking.

**Visualization** is the way of using tools or feasible techniques to represent the architectural structure through visual notations for helping practitioners have better insights into the system design and relationships within/between various components and modules. In design phases, visualizing architecture can help to avoid architecture violations; in the maintenance phase, it can be conducive to repairing various code and architecture anomalies. Hence, this type of measures could be classified into both preventive and remedial measures.

**Management optimization** denotes that optimizing the management methods of controlling or stopping the constant erosion tendency of the system architecture. For example, the author of [S73] claimed that creating a culture is valued to stop erosion with management support, and this culture is likely to have the characteristics, including emphasizing regular refactoring, clear assignment of responsibilities, sharing architectural knowledge, and frequent communication within development teams.

## 4.8 | RQ7: What are the difficulties when detecting, addressing, and preventing architecture erosion?

The difficulties refer to the obstacles of detecting, addressing, and preventing AEr during the development process. We collected and extracted the difficulties from the selected studies and classified them into three types. We list the main difficulties, types, and descriptions of the difficulties in Table 16.

**Detection of AEr**. Studies in this category can be classified into three types, as shown in Table 16. (a) Lack of dedicated techniques and tools/plug-ins for detecting AEr. For example, in [S24] and [S29], the authors mentioned that there was a lack of tools specifically geared towards addressing AEr. (b) Hard to establish a mapping relation. For instance, due to various reasons (e.g., lack of documentation and developer turnover), the authors of [S10] mentioned that it could be very difficult to know about the original design decisions for establishing mapping relations between source code and architectural elements (e.g., components and dependencies). When the correlation of this kind of understanding is missing, it will become the obstacle for identifying their impact on AEr. (c) Limitation of detection approaches. Due to the specialization of domains and software models, some of the existing approaches can hardly be reused in the typical software development. For example, the authors of [S13] claimed that the specification of consistency constraints depended on the syntax and definitions of specialized models, and it may need to

**TABLE 16** Classification for the difficulties of detecting, addressing, and preventing architecture erosion

| Category | Type | Description | Studies |
| --- | --- | --- | --- |
| Detection of AEr | Lack of dedicated techniques and tools | Due to a lack of efficient tools and techniques, it is very hard to manually detect AEr in software systems | [S10][S13][S24][S27][S29][S31][S35][S48][S57] |
| | Hard to establish mapping relations | Due to various factors, it is hard to establish mapping relations between source code and architectural elements (e.g., components and dependencies) | [S10][S26][S29][S36][S37][S39][S47][S51][S57] |
| | Limitation of detection approaches | Due to the specification of domains and software models, some methods can hardly be generalized to detect AEr in different development scenarios | [S9][S13][S14][S33][S35][S38][S48] |
| Handling of AEr | Hard to keep the architectural consistency | When the system evolves over time, it is hard to keep the implemented architecture align with the intended architecture | [S6][S18][S20][S45][S60][S71] |
| | Labor and time constraint | It is usually hard to manually keep the architecture consistent with source code and completely remove AEr due to the labor and time constraints | [S20][S45][S53][S54] |
| | Lack of broad understanding of software systems | Hard to get access to a broad understanding of the whole system for repairing AEr | [S60][S69] |
| Others difficulties | Verification | Hard to validate the effects of AEr repairing | [S46] |
| | Assessing the cost of stopping erosion | Hard to evaluate the cost to stop AEr in software systems | [S73] |

Abbreviation: AEr, architecture erosion.

repeatedly define all the models where AEr might occur. This kind of redundancy constrains the reusability of this type of methods and the extensive use for detecting AEr in industry.

**Handling of AEr**. Even if erosion is found in software systems, it is still a challenge to handle AEr. (d) Hard to keep the architectural consistency. For example, the authors of [S18] mentioned that keeping the intended architecture consistent with the implemented architecture is still very hard for software engineers in maintenance phases, because engineers must cope with obsolescence and maintenance when systems evolve. (e) Labor and time constraints. Manually ensuring the architecture keep consistent with the implementation of a system might be quite labor-intensive and time-consuming. For example, the authors of [S20] stated that manually ensuring the consistency between the implementation and the dependency model is a laborious, time-consuming, and error-prone task, even for smaller systems. (f) Lack of broad understanding. Lack of a broad and in-depth understanding of the whole system is also an obstacle for completely removing and repairing erosion (e.g., [S60] and [S69]). Hence, it is a challenge for practitioners to find optimal ways to fix the eroded architecture.

**Other difficulties**. (g) Verification. In most of the cases, maintainers are not the original developers; for example, the authors of [S46] mentioned that this situation increased the difficulties for optimizing the original modularization. Meanwhile, remodularizing the system structure might obtain new modularizations that are different from the original structure, and it is hard to validate the effect of such kind of erosion repairing. (h) Assessing the cost of stopping erosion. Besides, the authors of [S73] mentioned that it is hard to assess the cost of AEr to software systems and convey the cost to non-technical stakeholders.

## 4.9 | RQ8: What are the lessons learned about architecture erosion in software development?

The selected studies provide numerous and valuable lessons on the handling of AEr. The lessons discussed in most of the selected studies are about the experience on handling AEr. Generally, a lesson could be one or more sentences, or a paragraph; we collected more than 200 lessons learned from the selected studies. Moreover, we classified the collected lessons into five types below and provide representative examples of each type in Table 17.

**Tackling of AEr**. The lessons in this type contain the experience on how to tackle erosion in the software development life cycle. For example, in [S41] and [S51], the authors mentioned that blindly optimizing one symptom of AEr might make other symptoms worse. Therefore, it is necessary for developers to understand the relationships between AEr symptoms and maintenance activities.

**Manifestation of AEr**. This type of lessons refers to the manifestation of eroded architecture in software development, such as characteristics and impact on performance. For example, in [S1], the authors mentioned that not all deviations (e.g., short-term deviations only in few versions) in an architecture manifest erosion trend.

**Detection of AEr**. The lessons in this type are about the experience on detecting erosion in the software development process. For example, the authors of [S31] found that the drift process often intertwined with the erosion process, and detection of drift symptoms might help to detect erosion symptoms or vice versa.

**Understanding of AEr**. This type of lessons contains how practitioners understand AEr in software development, such as characteristics and attributes of eroded architecture in software development. For example, the authors of [S7] mentioned that almost all projects would suffer from erosion sooner or later unless taking some effort to overcome it.

**Prevention of AEr**. The lessons in this type include the findings about which means are effective in preventing AEr. For example, the authors of [S13] found that AEr might also occur in model-driven software development. Hence, formulating and regularly checking the architectural rules with the model-implementation process can help to prevent architecture from sliding into erosion during the software development life cycle.

## 5 | DISCUSSION

In this section, we analyze the results of each RQ (see Section 5.1) and discuss their implications for researchers (see Section 5.2.1) and practitioners (see Section 5.2.2), respectively. To demystify the AEr phenomenon, we first provide a conceptual model of AEr (see Figure 7) to present our findings and the relationships between various aspects (i.e., definitions, symptoms, reasons, and consequences) of AEr according to the results of this SMS in Section 4. This conceptual model acts as a panorama to understand the nature of the AEr phenomenon.

Regarding each aspect shown in Figure 7, we discuss their elements in Section 5.1. The conceptual model shows that the understanding of AEr (i.e., definition perspective) consists of four perspectives (see Section 4.2) and presents the types of AEr symptoms (see Section 4.3) that can be considered as the manifestation of AEr. On the other side, the AEr symptoms can directly or indirectly give rise to diverse consequences to software projects. Moreover, the conceptual model highlights that both technical and non-technical reasons can trigger the occurrence of AEr (see Section 4.4), as well as the negative effects caused by AEr (see Section 4.5).

TABLE 17  Lessons learned about architecture erosion in software development

| Lessons learned type | Description | Studies |
|---|---|---|
| Tackling of AEr | (1) Detecting and removing deviations might not efficiently mitigate erosion [S1].<br>(2) Dealing with AEr after it has happened is more difficult and costly, and AEr should be handled once it has been introduced in a system [S8]. | [S1][S7][S8][S11][S16][S17][S21][S27][S28][S31][S34][S37][S40][S41][S44][S45][S46][S47][S48][S51][S52][S53][S54][S55][S57][S60][S62][S69][S70] |
| Manifestation of AEr | (1) The authors found that the erosion process often intertwined with the drift process [S31].<br>(2) The authors learned that eroded architecture might not be observed by performance [S34]. | [S1][S7][S11][S17][S22][S27][S28][S30][S31][S34][S35][S41][S47][S49][S51][S53][S55][S56][S57][S62][S64][S68][S69][S73] |
| Detection of AEr | (1) Identifying the hotspots in the architecture can help to identify erosion [S1].<br>(2) The detection strategies of erosion must rely on certain captured architectural information [S27]. | [S1][S22][S27][S31][S33][S34][S35][S36][S49][S51][S52][S53][S57][S60][S63][S65][S66][S68][S69][S73] |
| Understanding of AEr | (1) The more the software system evolves, the greater possibility of erosion happens [S21].<br>(2) The authors mentioned that AEr cannot be avoided completely [S69]. | [S1][S7][S9][S13][S21][S33][S34][S35][S44][S45][S47][S53][S55][S56][S57][S60][S66][S69] |
| Prevention of AEr | (1) Establishing a culture and offering supportive management is crucial for preventing erosion [S7].<br>(2) The authors suggested that minimizing human interference in routine maintenance activities contributes to erosion prevention when employed self-adaptation strategies [S34]. | [S7][S11][S13][S34][S51][S53][S61][S73] |

Abbreviation: AEr, architecture erosion.

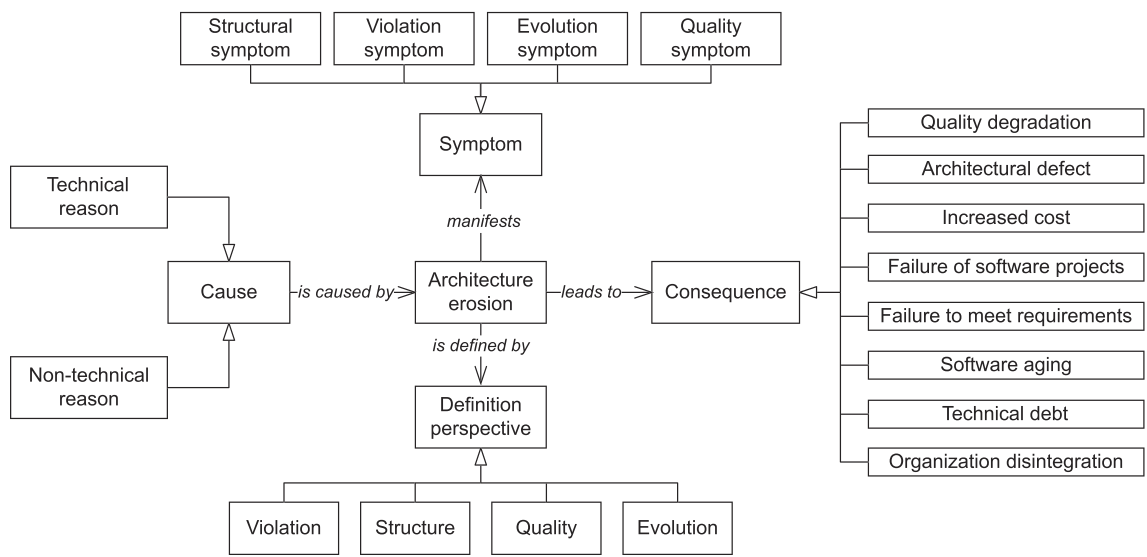

FIGURE 7  A conceptual model of architecture erosion according to the research question results

## 5.1 | Analysis of results

### 5.1.1 | Definition of architecture erosion

Section 4.2 reveals that the concept of AEr has been defined in different terms and perspectives. Table 8 shows that only 54.8% (40 out of 73) of the studies proposed a definition of the AEr phenomenon. According to the terms listed in Table 7, some of these terms are general concepts, which refer to the constant decay of the system structure (at the system level), such as software aging (e.g., [S5]), software erosion (e.g., [S41]), design rot (e.g., [S7]), system rot (e.g., [S72]), and modular/modularity deterioration (e.g., [S46]); certain terms are used to describe software anomalies at the code level, such as code decay and code rot (e.g., [S28]); some other terms focus on the AEr phenomenon at the architectural level, and the architectural problems are described as architecture degradation or architecture degeneration, which represents the introduction of architecturally relevant problems and continuous decline of architectural modularity (e.g., [S27], [S31], [S45], [S57], and [S67]). Moreover, according to the extracted definitions of AEr, we found that AEr, architecture decay, and architecture deterioration have the same meaning when describing the AEr phenomenon (e.g., [S15], [S16], [S17], and [S21]), which refers to the architectural deviation between the intended and implemented architectures, and both the intended and implemented architectures may change with the changing demands of stakeholders.

Besides, certain definitions of AEr are not included in this SMS, because the publication years of these studies are not within the time period of our SMS. For example, Jaktman et al.[52] defined AEr as "*structure of a software architecture to be eroded when the software becomes resistant to change or software changes become risky and time consuming*," and this definition focuses on the structure perspective (see Table 8); Wang et al.[53] defined that "*architecture erosion is a phenomenon that occurs when architecture quality is decreased with software evolution*," and this definition focuses on the quality perspective (see Table 8). The two definitions above can also support that AEr is more than a simple phenomenon about the violations of intended architecture, and the relationships between the four perspectives are worth further investigation. Furthermore, the AEr definitions in four perspectives imply that AEr occurred at different levels of abstraction, and this finding also shows that AEr is a multifaceted phenomenon. The results of our recent study from the practitioners' perspective[14] also indicate that AEr is a multifaceted phenomenon and practitioners describe this phenomenon from four perspectives: structure, quality, maintenance, and evolution. Interestingly, we noticed that there is only a slight difference in the perspectives of AEr between academia and industry. Compared with the violation perspective identified in this SMS, practitioners care more about the maintenance perspective of AEr. One potential reason could be that the maintainers may not be the original architects in practice; thus, practitioners discuss more about the maintainability issues caused by AEr, while academia might focus more on the nature of AEr. Considering this, we provide a refined definition of AEr according to the results (i.e., the four perspectives) of our SMS: *AEr happens when the implemented architecture violates the intended architecture with flawed internal structure or when architecture becomes resistant to change.*

> **Finding 1**: *AEr is a multifaceted phenomenon, which is often described from four perspectives: violation, structure, quality, and evolution.*
>
> **Finding 2**: *We provide a refined definition of AEr according to the results (i.e., the four perspectives) of our SMS: AEr happens when the implemented architecture violates the intended architecture with flawed internal structure or when architecture becomes resistant to change.*

### 5.1.2 | Symptoms of architecture erosion

Table 9 shows that the symptoms of AEr can be mapped into the perspectives of AEr phenomenon (see Section 4.2). We found that 32.9% (24 out of 73) of the selected studies mention the symptoms of AEr, and nearly half of them (41.7%, 10 out of 24) regard violations of design decisions, principles (e.g., abstraction and encapsulation), or constraints as the symptoms of AEr. Moreover, most of the symptoms (70.8%, 17 out of 24) are related to structural anomalies, especially some architectural smells (e.g., undesired dependencies).

According to Table 9, we observed that structural symptoms receive more attention among the majority of symptoms. One potential reason could be that it is relatively easy to detect structural issues by employing various tools or visualizing the dependency relationships of components. Besides, compared with individual structural symptoms, the agglomerations of structural symptoms are even stronger indicators of AEr. For example, Oizumi et al.[54,55] studied the impact of code anomaly agglomeration (i.e., a group of interrelated code anomalies) on software architecture and the results show that more than 70% of architectural problems are related to code anomaly agglomerations. Their studies confirm that code anomaly agglomerations are better than individual code anomalies to indicate the presence of architectural problems. Besides, microservices architecture (MSA), as a popular architectural style used to increase resilience to AEr,[56] can also show certain AEr symptoms (such as architectural smells). For example, Mumtaz et al.[57] presented several service-oriented and service-specific architectural smells (e.g., shared libraries and missing package abstractness) in MSA, which are a typical type of structural symptoms of AEr (see Table 9). Considering the popularity of using MSA in industry in recent years, the potential shortcomings and risks of MSA related to AEr need to be further investigated.

In addition, we found that it is a challenge to adequately quantify various AEr symptoms, though several studies (e.g., [S16], [S41], and [S26]) attempted to propose metrics and tools for detecting AEr symptoms. For example, ARCADE mentioned in [S16] is a tool used to analyze and quantify different structural symptoms (e.g., architectural smells and dependency information) for given systems, which is employed to identify and recover the eroded architecture (e.g., references[58,59]). A potential drawback of entirely relying on structural symptoms might ignore other AEr symptoms (e.g., evolution symptoms). Quantifying the AEr symptoms can help to recognize AEr and its degree in software development. Although a wide variety of studies mentioned diverse AEr symptoms, there is a lack of diversity about AEr symptoms in the empirical validations. For example, we did not find many in-depth validations of which symptoms are strong indicators of AEr and how to detect those symptoms.

> **Finding 3**: AEr symptoms can be classified into four categories: structural symptom, violation symptom, evolution symptom, and quality symptom, where structural symptoms receive the most attention.

### 5.1.3 | Reasons of architecture erosion

Every architecture will undergo erosion sooner or later as long as the evolution happens,[60] which is a process of increasing entropy. In a broad sense, AEr does not arise spontaneously and the root reason of AEr comes about through *change*. According to Tables 10 and 11, we observed that the reasons of AEr are more than from technical factors, but also related to non-technical factors. The findings show that *architecture violation* (24.7%, 18 out of 73 studies), *evolution issue* (23.3%, 17 out of 73 studies), and *technical debt* (17.8%, 13 out of 73 studies) are the three main technical reasons of AEr, and *knowledge vaporization* (15.1%, 11 out of 73 studies) is the most frequently mentioned non-technical reason. One possible reason is that the maintenance and evolution phases account for a large part of the life cycle of software development, and these frequently mentioned reasons are relatively easy to be involved in various artifacts and software development activities. *Requirement issue* (15.1%, 11 out of 73 studies) is a common reason of AEr related to both technical and non-technical reasons, because requirements (including functional and nonfunctional requirements) have a significant impact on system design and evolution, and the requirements–design gap can affect different development activities during the software development life cycle.

Additionally, the reasons of AEr listed in Table 10 may have interactive relationships. For example, *increasing complexity* can lead to *understanding issue* (e.g., [S34]), while *understanding issue* may be caused by other reasons, such as *architecture violation* (e.g., [S45]). *Technical debt* might give rise to *suboptimal design decisions*, but not all of the *suboptimal design decisions* are derived from *technical debt* (e.g., specific design constraints with low extensibility). Thus, the potential relationships between the reasons are also worthy of further exploration. We believe that our work can serve as a good starting point for future research on exploring the relationships between the collected reasons.

> **Finding 4**: AEr does not arise spontaneously and the root reason of AEr comes about through *change*. AEr is caused by both technical factors (e.g., architecture violation, evolution issue, and technical debt) and non-technical factors (e.g., knowledge vaporization), which are intertwined.

### 5.1.4 | Consequences of architecture erosion

As shown in Table 12, we mapped the consequences extracted from the selected studies into eight categories; 50.7% (37 out of 73) of the studies mention or investigate the impact of eroded architecture on software systems, where most of them (83.8%, 31 out of 37 studies) are related to quality degradation issues. The priority of tackling the consequences relies on the severity and degree of impact on different software Quality Attributes (QAs) (e.g., maintainability, evolvability, and extensibility) and the significance of these QAs to the companies. For instance, Mozilla web browser,[61] an application from Netscape comprised over 2,000,000 source lines of code, is a frequently mentioned example of AEr in industry. Netscape engineers took 6 months to conclude that the original architecture layers of this application were eroded and irreparable. However, they spent 2 years redeveloping Mozilla web browser and re-architecting the source code and dependencies. For small companies, it might lead to *failure of software projects*, while it might bring about *increased cost* to redevelop this application for big companies like Netscape.

Not surprisingly, the consequences might damage architectural structures and generate software defects. Ultimately, massive labor and time cost have to be invested into the projects for incorporating the new requirements and restructuring the source code, and the worst situation is to rebuild a system from scratch. The results of RQ4 can shed light on the significance of AEr in software development. Knowing about various consequences can draw more attention to AEr phenomenon and further warn researchers and practitioners to avoid the potential consequences of AEr. Notably, the findings indicate that technical debt might not only be one of the AEr reasons but also one of the AEr consequences. It is worth noting that correlation does not imply causation and the accumulated unpaid technical debt might give rise to AEr with high probability. For example, Mo et al. [S49] mentioned that systems with AEr happened can incur technical debt where short-term compromises lead to significant long-term problems (e.g., fixing bugs or adding new features). The findings imply that technical debt might have a feedback loop with negative consequences, including decreasing reliability and increasing complexity. This observation suggests that more empirical work should be performed to investigate the relationships between technical debt and AEr.

> *Finding 5*: AEr can lead to various consequences, such as damaging architectural structures and generating software defects. The priority of tackling the consequences of AEr often relies on the severity and degree of impact on different software quality attributes.
> *Finding 6*: Technical debt might not only be one of the AEr reasons but also one of the AEr consequences, forming a vicious circle between the two phenomena.

### 5.1.5 | Approaches and tools for architecture erosion detection

We found that 54.8% (40 out of 73) of the selected studies proposed or employed approaches for detecting AEr and we classified these approaches into four categories (see Table 13). The results of RQ5 reveal that the research effort on AEr detection has mainly focused on the consistency-based approaches (62.5%, 25 out of 40 studies), where the most frequently employed evaluation-based approach is Architecture Conformance Checking (ACC). One potential reason is that ACC is straightforward to assist stakeholders in checking whether the implemented architecture complies with the intended architecture. Moreover, our results show that the evolution-based approaches (22.5%, 9 out of 40 studies) and defect-based approaches (15.0%, 6 out of 40 studies) receive similar concerns from the literature. One reason may be that the diversity of source code analysis tools makes the two types of approaches more applicable to different software projects.

The findings also show that 39.7% (29 out of 73) of the selected studies mentioned relevant tools that can be employed in detecting AEr (see Table 14), where the majority of the tools (57.1%, 20 out of 35 tools) used to detect AEr are designed based on ACC. Additionally, we observed that many tools employed to detect AEr are implemented to support the approaches in Table 13 besides ACC, which demonstrates the necessity of close collaboration between academia and industry. Moreover, we noticed that half of the tools are commercial tools and more than half of the tools only support one specific programming language, which could be an obstacle for the tools to be widely employed in industry, especially when multiprogramming-language development is booming.[62] Although some tools are not specifically devoted to AEr detection, to some extent, these tools can provide support for detecting AEr symptoms, which would be beneficial to identify architectural anomalies at the early stage of AEr occurrence during the software development life cycle.

In our recent work,[14] we found that developers claimed that there are no dedicated tools for detecting AEr. However, according to the results in this SMS, for example, the Axivion tool can help to detect and stop AEr. One potential reason is that a gap still exists on AEr between research and practice, and developers might not be familiar with the existing approaches and tools in the literature. Therefore, AEr detection approaches and tools can be empirically evaluated through case studies. For instance, it is interesting to see some approaches and tools proposed in the literature, but there is limited empirical evidence regarding their effectiveness and productivity. Hence, we encourage more collaborations between academia and industry in the area of AEr detection and advocate more empirical studies on the validation of existing approaches and tools.

> *Finding 7*: The majority of the approaches and tools used to detect AEr are mainly focused on architectural consistency. The diversity of the existing approaches and tools should be increased, because more than half of the tools only support one specific programming language.
> *Finding 8*: A gap still exists on AEr between research and practice. More dedicated approaches and tools should be designed to detect AEr and empirical studies should be conducted to evaluate the existing approaches and tools.

### 5.1.6 | Measures for addressing architecture erosion

We identified and collected the measures for addressing AEr from the selected studies and classified the measures into three categories (see Table 15). Table 15 shows that most of the preventive measures are employed to prevent AEr by checking, monitoring, and evaluating system architecture. One reason is that establishing and checking the consistency between the specifications and implementation (e.g., using formal methods and traceability mechanisms) can effectively eliminate the possibility of deviating the implemented architecture from the intended architecture. In comparison with preventive measures, remedial measures are maintenance-oriented measures and closely related to reverse engineering and re-engineering techniques, such as architecture recovery. To address AEr in software systems, engineers need to understand the architectural structure (e.g., the layered pattern) and reasons about the changes through the recovered architecture (e.g., restoring architecture views from source code). According to Table 15, remedial measures are not as widely employed in practice compared with preventive measures. The reason could be that the extent of architecture recovery is largely dependent on code quality (e.g., code readability) and the availability of architecture knowledge. In addition, if the system has been severely eroded, then maintenance may not be a cost-effective choice; consequently, discarding the eroded system and building a new system from scratch would be a feasible option in terms of return on investment.

Nevertheless, prevention is better than cure. In recent years, Microservices Architecture (MSA), as a cloud-native architecture, has grown in popularity in industry due to its benefits of flexibility, loose coupling, and scalability. To some extent, decomposing the original monolithic architecture into microservices can improve architecture extensibility and increase its resilience to AEr. Moreover, with the rising popularity of MSA, there is a growing number of companies choosing to migrate their existing monolithic applications to MSA through decomposition and service interactions. Migrating a monolithic architecture to a cloud-native architecture like MSA can help engineers maintain large software systems, through increasing flexibility to adopt new technologies, reducing the time-to-market, and managing independent resources for diverse components.[63] For example, Chen[56] reported their practices and experience when they moved their eroded applications to an MSA, and they observed increased deployability, modifiability, and resilience to AEr. The reason is that MSA can create physical boundaries between microservices, because each microservice has its own codebase, development team, and runs in its own containers. Thus, MSA provides better protection against the temptation of breaking the boundaries between services (a typical violation symptom of AEr, see Table 9). Therefore, architects can choose well-accepted architecture patterns (e.g., MSA) to gain control of the systems when facing increasing complexity, as well as adhere to design principles that allow the decomposition of a complex system into more understandable chunks.[46] Note that, although MSA provides a promising way to deal with complicated architectural issues, MSA is definitely not a silver bullet and it can raise new architectural problems. For example, an experience report from Balalaie et al.[63] indicates potential risks that might be related to AEr; for example, microservices implemented based on different programming languages could lead to major issues, rendering the system unmaintainable.

> **Finding 9**: Measures used to address AEr include preventive measures (e.g., architecture conformance checking and architecture monitoring and evaluation) and remedial measures (e.g., architecture maintenance and restoration).

### 5.1.7 | Difficulties on handling architecture erosion

As shown in Table 16, we classified the difficulties in detecting, addressing, and preventing AEr into three categories. Most of the difficulties presented in Section 4.8 are still challenges in the current research and practice of software engineering, but a few difficulties have been partly addressed by the measures presented in Section 4.7, such as establishing the mapping relationships between design and architectural elements. For example, there are still no effective mechanisms to deal with the consistency between System-of-System (SoS) architectural instance models and SoS abstract architecture models (e.g., [S29]). Besides, architects and developers need suitable tool support for AEr detection, and the lack of dedicated tools raises difficulties for detecting and addressing AEr. For example, Python has become a popular programming language in recent years,[1] while we found that more than half of the tools (see Section 4.6.2) only support one specific programming language (e.g., Java) and only a few tools support Python projects. Additionally, the results reflect the significance of understanding and handling AEr. It is critical to raise the awareness of the AEr risks in software systems and identify potential AEr symptoms, because having a holistic understanding of the whole system is the prerequisite for mitigating the negative consequences of AEr.

> **Finding 10**: The difficulties of detecting, addressing, and preventing AEr are mainly derived from the detection and handling of AEr, while we still lack effective mechanisms (e.g., dedicated approaches and tools) to handle AEr in various software systems (e.g., SoS and MSA).

---

[1]https://insights.stackoverflow.com/survey/2020.

### 5.1.8 | Lessons learned on handling architecture erosion

As mentioned in Section 4.9, the lessons refer to the experience on handling AEr. The lessons learned on handling AEr are classified into five types, and we provide two examples for each type of lessons learned (see Table 17). The findings show that the majority (39.7%, 29 out of 73) of the selected studies mentioned the lessons about the tackling of AEr, followed by lessons about the manifestation of AEr (32.9%, 24 out of 73 studies) and detection of AEr (27.4%, 20 out of 73 studies). One reason is that maintenance and evolution phases account for a large part of the software development life cycle, and AEr phenomenon is most likely to be observed by maintainers and testers during the maintenance and evolution phases. Moreover, we observed that only 11.0% (8 out of 73) of the studies discuss the lessons about AEr prevention. One potential reason is that AEr prevention is relatively difficult and not always feasible. For example, the authors of [S34] mentioned that "*preventing erosion completely is a difficult task and may not be feasible*", and the authors of [S69] mentioned that "*in general architecture erosion cannot be avoided completely*". Another reason is that there are various causes related to AEr (see Section 4.4) that it is nearly impossible to completely prevent the reasons from happening, especially for complex software systems with millions of lines of code. To some extent, AEr prevention means that taking feasible measures to extend the software system life cycle as long as possible. Additionally, we suggest that stakeholders should raise awareness of AEr and create a culture of organizational support, and practitioners should be encouraged to fight for AEr.

> *Finding 11*: More than 200 lessons are collected from the selected studies that are concentrated on five types: tackling, manifestation, detection, understanding, and prevention of AEr; the majority of these lessons concern tackling AEr.

## 5.2 | Implications

This SMS provides a comprehensive insight into the AEr-relevant studies, and the results of the RQs provide significant implications for both researchers and practitioners. In this section, we discuss the implications of the findings and highlight the promising research directions on AEr.

### 5.2.1 | Implications for researchers

As shown in Figure 4, we found that most (82.2%, 60 out of 73) of the selected studies are solely from academia, and we encourage researchers to seek more collaborations with industrial partners to fill the gap between academia and industry and solve the existing challenges.

According to the results of RQ1 (see Section 4.2.1) and their analysis (see Section 5.1.1) about the AEr definitions, we highly recommend that both researchers and practitioners can **describe and define the term when they refer to AEr phenomenon and use the common term "architecture erosion"** consistently for minimizing ambiguities and misunderstanding. In addition, the results of RQ1 (see Section 4.2.2) and their analysis (see Section 5.1.1) also indicate that **AEr is a multifaceted phenomenon** and it not only manifests through architectural violations and structural issues but also affects the quality and evolution of software systems. Future research should consider the four perspectives of AEr when evaluating architecture and analyzing architecture evolution. For example, different metrics-based evaluations can be conducted from the four perspectives when software engineers are engaged in maintenance and evolution activities.

As discussed in Section 5.1.2 (the analysis of the results of RQ2), although a wide variety of studies mentioned diverse AEr symptoms, we notice that there is **a lack of empirical validation of these AEr symptoms**. Researchers can investigate which AEr symptoms are strong indicators of AEr and how to detect and further quantify the symptoms. For example, researchers can attempt to establish a comprehensive evaluation method by either employing the existing or proposing new approaches and tools to evaluate and quantify the structural symptoms (e.g., architectural smells).

The results of RQ3 (see Section 4.4) show that **the occurrence of AEr in software systems may derive from various reasons**, and researchers can empirically investigate the potential relationships between the reasons of AEr. Additionally, we notice that technical debt might be related to part of the technical and non-technical reasons of AEr. We believe that it is valuable to investigate when and how technical debt can induce AEr.

The results of RQ4 (see Section 4.5) and their analysis (see Section 5.1.4) show that **the prevalence of AEr leads to varying degrees of impact on software systems, and not all types of consequences have received the same attention**. As shown in Table 12, researchers are more likely to notice the quality attributes degradation when AEr happens. It would be interesting to empirically study the measures and costs taken to handle the eroded architecture. For example, researchers can conduct case studies with industrial partners to investigate the financial loss of different consequences identified in this SMS brought into the software projects suffered from AEr.

As discussed in Section 5.1.5 (the analysis of RQ5 in Section 4.6), although our recent work[14] investigated part of the causes and consequences of AEr from the practitioners' perspective, there remains **a dearth of empirical case studies on exploring the reasons and consequences of AEr** in a real-life industrial setting. Besides, comparative studies about different **approaches for tackling AEr are deficient** in the field of software architecture. Therefore, further research can focus on exploring the potential reasons and consequences of AEr, as well as the effectiveness of relevant approaches for tackling AEr using industrial cases.

Although some approaches and tools (the results of RQ5; see Section 4.6) can be useful to detect AEr (e.g., Axivion, Structure101, and Lattix), there exists a demand to evaluate the performance of different approaches and tools and verify the capability (e.g., merits and demerits) of these approaches and tools. We encourage researchers to conduct empirical studies for **evaluating these AEr detection approaches and tools**. Besides, researchers can evaluate the approaches and tools reported in Section 4.6 and explore their scope, characteristics, and metrics for providing a solid foundation on designing dedicated tools.

As discussed in Section 5.1.6 (the analysis of the results of RQ6), there are **scarce research on the measures for addressing AEr in emerging systems, architecture styles, and development methods**, such as SoS, MSA, and DevOps, which is an interesting and meaningful research field to be explored. For example, researchers can pay more attention to the potential risk of AEr (e.g., erosion symptoms and the gap between designed and implemented architectures) in MSA-based systems.

From the identified difficulties (the results of RQ7 in Section 4.8 and their analysis in Section 5.1.7) and the lessons learned on handling AEr (the results of RQ8 in Section 4.9 and their analysis in Section 5.1.8), we observed that the results provide the **challenges in this area that would inspire more academic and industrial collaboration and promising measures for the identification, handling, and prevention of AEr**. For example, automatically detecting the AEr symptoms by leveraging machine learning techniques can be an effective and efficient way compared with manual analysis of components and source code.

## 5.2.2 | Implications for practitioners

As discussed in Section 5.1.1 (the analysis of the results of RQ1), practitioners are encouraged to reach an agreement on the **understanding of AEr phenomenon for reducing the ambiguity of the AEr concept**. For example, we recommend practitioners to use the common term "architecture erosion" to refer to the AEr phenomenon, which helps to minimize the misunderstanding and confusion for reaching a common ground on the description of the AEr phenomenon.

**There are diverse technical reasons and non-technical reasons that can give rise to AEr** (see Tables 10 and 11, the results of RQ3). It is critical for practitioners to realize that certain technical reasons (e.g., technical debt) and non-technical reasons (e.g., knowledge vaporization) of AEr may not have a noticeable effect on system quality in a short time, but they can negatively impact architectural understanding and integrity in the long term, thereby leading to AEr. For example, practitioners can attach importance to the non-technical reasons of AEr and cultivate a culture of AEr prevention, such as exposing and discussing architectural changes, improving organization management skills (especially about the training and education of developers to help them be familiar with system architecture).

The analysis of the results of RQ4 in Section 5.1.4 indicate that practitioners need to **raise awareness of the grave consequences of AEr and request actions at the management level**, in order to obtain a high priority to tackle AEr-related issues and reduce the risk of architecture sliding into erosion and failure. To prevent and repair AEr to some extent, we suggest that practitioners should build the mechanisms that support recording and tracing architectural knowledge (e.g., decisions and assumptions) to source code and architectural changes and regularly conduct architecture assessment (e.g., architecture conformance checking).

Practitioners should also pay attention to the state of the art of academic research results, because we noticed that **there is a gap between research and practice** (see the analysis of the results of RQ5 in Section 5.1.5). Practitioners are encouraged to apply and adapt the proposed approaches and tools according to their needs and project contexts, which can help to identify not only the limitation of those approaches and tools but also the real needs from practitioners for detecting AEr in practice.

It is critical to conduct early diagnoses of AEr in software systems. **The approaches and tools for detecting AEr are still scarce and have some limitations** (e.g., domain/language-specific features). The results of RQ5 (see Section 4.6) imply that there exists a demand for dedicated approaches and tools to support AEr detection in practice. Therefore, practitioners are encouraged to collaborate with researchers and employ dedicated approaches and tools for detecting AEr. Such dedicated tools can better support multiple programming languages, detection of the AEr symptoms, and the quantitative visualization of the trend of AEr.

**Some measures used to address AEr are specific to the contexts in the studies** (see the analysis of the results of RQ6 in Section 4.7). When it comes to AEr prevention and remediation, practitioners need to first understand the corresponding contexts of employing the measures and then adapt the measures according to the specific situations.

It is never too late to fight against AEr. The difficulties (see the analysis of the results of RQ7 in Section 4.8) and valuable lessons (see the analysis of the results of RQ8 in Section 4.9) we collected can be beneficial for practitioners to better cope with AEr during the development life cycle, avoid the pitfalls of AEr, and further explore the benefits and limitations of applying the measures for addressing AEr in practice. Besides, further industrial evidence is required to validate the benefits of the measures as well as the potential pitfalls.

To sum up, the findings of this mapping study can help practitioners to better understand the practical impact of the AEr phenomenon, supplementing our previous industrial survey on AEr[14] and empirical study on AEr symptoms identified by code review in OSS projects.[64] Specifically, Findings 1, 2, and 3 provide a comprehensive understanding of the AEr phenomenon; Findings 4, 5, and 6 shed light on the potential reasons and consequences of AEr; Findings 7 and 8 show the increasing demand for dedicated AEr detection tools; and Findings 9, 10, and 11 highlight the importance of tackling AEr in practice.

# 6 | THREATS TO VALIDITY

In this section, we discuss the potential threats to the validity of our SMS, as well as the measures that we took to mitigate the threats according to the guidelines in previous works.[65,66]

## 6.1 | Construct validity

Construct validity concerns whether the theoretical and conceptual constructs are correctly interpreted and measured. In this SMS, the main threats to construct validity include study search and selection:

**Study search**. There may be relevant studies that were omitted, which will affect the completeness of the retrieved results. To mitigate this threat, we searched seven popular electronic databases (see Table 2) that publish software engineering research, and we also conducted a manual search from well-known venues (including conferences, journals, and workshops) closely related to the topic of our SMS, that is, AEr. Before the formal search, a pilot search was performed to enhance the suitability (e.g., the search terms and time period) and quality of this SMS. Additionally, to ensure the completeness of the study search, we employed the "snowballing" technique[31] to include any potentially relevant studies.

**Study selection**. Whether or not to include relevant studies mainly depends on the understanding of the researchers who were involved in the study search and selection, and this inevitably introduced personal bias due to the limitation of personal knowledge. To minimize the selection bias, (1) we selected 100 papers as a sample, which were selected by the first two authors independently in three rounds, to measure their inter-rater agreement on study selection; (2) we also defined a set of inclusion and exclusion criteria regarding the study selection and discussed among all the authors about the uncertain studies for reaching an agreement and reducing the risk that relevant studies were omitted.

## 6.2 | Internal validity

Internal validity pertains to the study design that has a potential impact on the results. In this SMS, the main threats to internal validity are concerned with the extracted data and the synthesis of the results:

**Data extraction**. One potential threat is about the quality of the extracted data, which might have a negative impact on the data synthesis and classification results. Several measures were taken to mitigate the bias of researchers who conducted data extraction. First, to ensure the consistency of data extraction results, we discussed and formulated the data items by all the authors to reach a consensus on the content of data to be extracted. Moreover, before the formal data extraction, a trial data extraction with five selected studies was performed by the first author and checked by the second and third authors. Furthermore, we selected another five papers from the selected results to conduct a data extraction by the first and second authors, independently. Any disagreement was discussed and resolved together for reaching an agreement about the understanding of the data items (specified in Table 4). Furthermore, the data extraction results from the first author were checked by the second and third authors according to the description of each data item.

**Data synthesis**. The quality of data synthesis may affect the correctness of the answers to the eight RQs (see Section 3.1). Different researchers may have their own understanding on the extracted data, for example, the classifications of extracted data. To minimize the personal bias, we conducted continuous discussions about the divergent classifications and any conflicts were discussed until an agreement was reached. Besides, during the data synthesis process, we excluded the extracted data if we could not find any valuable information.

## 6.3 | External validity

External validity concerns the extent to which the findings of a study can be generalized. This SMS provides an overview of the state of the art of the studies on AEr phenomenon in software development. Although we took measures to increase the completeness of our study search, there are still limitations and we may miss relevant studies that investigate AEr but are not covered by this SMS. To alleviate the threats to external validity, we provided the search protocol of this SMS (see Section 3) that rigorously specifies the execution process of this SMS, and we searched and selected peer-reviewed studies using both popular electronic databases and relevant target venues (i.e., the manual search).

## 6.4 | Reliability

Reliability refers to whether the study produces the same results when other researchers replicate it. This threat is related to several factors, such as missing studies, incomplete data extraction, and improper categorization of the extracted data. We only searched and selected relevant studies written in English according to the inclusion and exclusion criteria. Nevertheless, relevant studies might still be omitted for various reasons, such as different terms used to describe the phenomenon of AEr and incompleteness of the selected venues. Additionally, a replication package of this SMS has been made available online[35] to improve its replicability. With these measures, we strove to ensure that this SMS can be repeated by following the procedure in Section 3.

## 7 | RELATED WORK

To the best of our knowledge, there are no SMSs or SLRs that specifically focus on the AEr phenomenon. However, there are several secondary studies that are close to the topic of our SMS (i.e., AEr). Therefore, we provide a brief overview of these secondary studies in this section.

Baabad et al.[67] conducted an SLR with 73 primary studies focusing on the architecture degradation in OSS projects. They performed a coarse-grained SLR that mainly focuses on the reasons and solutions of architectural degradation problems in OSS projects. Their results show that architectural degradation problems, including identifying, addressing, avoiding, and predicting architectural degradation within OSS projects, are still open research issues. Compared with their SLR, our SMS has a broader scope on the AEr phenomenon. We analyzed the difference between various terms of the AEr phenomenon, provided a refined definition of AEr, and systematically analyzed and categorized the reasons, consequences, measures, difficulties, and lessons learned of AEr.

Neri et al.[68] recently conducted a multivocal review on the architectural smells possibly violating the design principles of MSA. The authors identified the design principles of MSA from 54 peer-reviewed papers and gray literature published before the end of January 2019. This multivocal review focuses on the architectural smells on MSA and why the smells violate the design principles, as well as the refactorings used to fix the smells. Besker et al.[69] conducted an SLR on architectural technical debt. They proposed a unified model of architectural technical debt that provides a new and comprehensive interpretation of architectural technical debt, because negative effects (e.g., AEr) might be caused by architectural technical debt, which is also the result of this SMS (see Section 4.4). Bandi et al.[18] conducted an SMS with 30 studies investigating the techniques and metrics that have been empirically evaluated for addressing code decay. Their SMS mainly covers the identification and minimization approaches for code decay. Herold et al.[70] reported the preliminary results of a literature review about architecture degradation and consistency checking. Their results show that empirical evaluation on this field is still missing and experiments and surveys should be complemented for assessing the impact of architectural degradation in practice.

The aforementioned secondary studies have a different focus, and our mapping study is the only one that focuses on the AEr phenomenon. To be more specific, we systematically analyzed and categorized the concept, symptoms, reasons, consequences, detecting, handling, and lessons learned of AEr through this SMS.

## 8 | CONCLUSIONS

Through this SMS, we investigated the understanding, reasons, and consequences of AEr, the available approaches and tools for detecting AEr and the measures for tackling AEr, as well as the difficulties and lessons learned from the selected studies. We searched the papers in seven electric databases and manually checked the papers in 22 related venues published between January 2006 and May 2019 on the topic of AEr in software development and finally selected 73 primary studies for further data extraction and analysis. Based on the extracted data, we got a comprehensive understanding of the concept of AEr and an overview of various aspects of AEr. The main points of this SMS are summarized as follows:

(1) Most (82.2%) of the selected studies are from academia, and the studies from conferences (58.9%) and journals (27.4%) make up the large majority of the selected papers. The tendency in Figure 5 shows that AEr-related studies received a fair amount of attention in the last decade.

(2) Among the different terms used to describe the AEr phenomenon, "architecture erosion" is the most frequently used term followed by "architecture decay." Both practitioners and researchers should clearly define the terms when they mention the AEr phenomenon and better use the common term "architecture erosion" to refer to the phenomenon.

(3) AEr manifests not only through architectural violations and structural issues but also causing problems in software quality and during the evolution of software systems. The four perspectives are worthy of investigation in both research and practice.

(4) The AEr symptoms can be classified into four categories according to the four perspectives of AEr, and structural and violation symptoms are the most common symptoms. Additionally, AEr exists not only in maintenance and evolution phases but may also exist in the design and implementation phases.

(5) Apart from technical reasons, non-technical reasons are also significant factors that lead to AEr, which indicates that non-technical reasons (e.g., management and organization issues) cannot be ignored, and AEr can exist in different phases of software development and interacts with various stakeholders.

(6) Most of the studies deem that AEr has a negative impact on QAs of software systems (e.g., maintainability and evolvability), while other consequences can also generate varying degrees of impact on systems. Practitioners should raise their awareness of the grave consequences of AEr and request actions at the management level, in order to obtain a high priority to tackle AEr-related issues and reduce the risk of architecture sliding into erosion and failure.

(7) The 19 approaches for detecting AEr are classified into four categories, in which consistency-based approaches and evolution-based approaches are the most frequently mentioned approaches. Besides, 35 tools used to detect AEr were classified into four categories according to their distinctive purposes (e.g., checking architecture conformance).

(8) The identified approaches, tools, and measures for detecting and addressing AEr are applicable in specific contexts (e.g., constraints or preconditions). Practitioners need to first understand the corresponding contexts of employing the measures and then adapt the measures according to the specific situations.

(9) The difficulties in detecting and tackling AEr mainly derive from the domain and language constraints, absence of dedicated tools, and establishing mapping and alignment between the intended and implemented architectures. These difficulties should receive more attention in future studies.

(10) The experience and lessons learned collected from the primary studies about AEr were classified into five categories, which can provide evidence for researchers to future investigation on AEr and help practitioners avoid pitfalls when addressing AEr.

We believe that the results of this SMS could benefit the researchers and practitioners to better understand the nature of AEr, as well as its underlying reasons and consequences, and the findings can provide clues for future research in this area (as discussed in Section 5). We encourage more collaboration between researchers and practitioners to close the gap between academia and industry and address the existing challenges. Additionally, we plan to conduct studies on detecting and handling AEr, including the detection of AEr symptoms and the approaches and tools to support the detection at various granularity levels.

### ACKNOWLEDGMENTS

This work is partially sponsored by the Natural Science Foundation of China (NSFC) under Grant No. 62172311 and ITEA3 and RVO under Grant Agreement No. 17038 VISDOM (https://visdom-project.github.io/website/). The authors gratefully acknowledge the financial support from the China Scholarship Council.

### ORCID

*Peng Liang*  https://orcid.org/0000-0002-2056-5346

### REFERENCES

1. Perry DE, Wolf AL. Foundations for the study of software architecture. *ACM SIGSOFT Softw Eng Notes*. 1992;17(4):40-52.
2. Koziolek H, Domis D, Goldschmidt T, Vorst P. Measur architecture sustainability. *IEEE Softw*. 2013;30(6):54-62.
3. De Silva L, Balasubramaniam D. Controlling software architecture erosion: a survey. *J Syst Softw*. 2012;85(1):132-151.
4. Macia I, Arcoverde R, Garcia A, Chavez C, von Staa A. On the relevance of code anomalies for identifying architecture degradation symptoms. In: Proceedings of the 16th European Conference on Software Maintenance and Reengineering (CSMR). IEEE; 2012; Szeged, Hungary:277-286.
5. Dalgarno M. When good architecture goes bad. *Methods Tools*. 2009;17(1):27-34.
6. Baum D, Dietrich J, Anslow C, Müller R. Visualizing design erosion: how big balls of mud are made. In: Proceedings of the 6th IEEE Working Conference on Software Visualization (VISSOFT). IEEE; 2018; Madrid, Spain:122-126.
7. Izurieta C, Bieman JM. A multiple case study of design pattern decay, grime, and rot in evolving software systems. *Softw Qual J*. 2013;21(2):289-323.

8. Li Z, Long J. A case study of measuring degeneration of software architectures from a defect perspective. In: Proceedings of the 18th Asia-Pacific Software Engineering Conference (APSEC). IEEE; 2011; Ho Chi Minh, Vietnam:242-249.
9. Hassaine S, Guéhéneuc Y-G, Hamel S, Antoniol G. ADvISE: Architectural decay in software evolution. In: Proceedings of the 16th European Conference on Software Maintenance and Reengineering (CSMR). IEEE; 2012; Szeged, Hungary:267-276.
10. Bandi A, Allen EB, Williams BJ. Assessing code decay: a data-driven approach. In: Proceedings of the 24th International Conference on Software Engineering and Data Engineering (SEDE). ISCA; 2015; San Diego, CA, USA:1-8.
11. Rama GM. A desiderata for refactoring-based software modularity improvement. In: Proceeding of the 3rd Annual India Software Engineering Conference (ISEC). ACM; 2010; Mysore, India:93-102.
12. Petersen K, Vakkalanka S, Kuzniarz L. Guidelines for conducting systematic mapping studies in software engineering: An update. *Inform Softw Technol*. 2015;64:1-18.
13. Kitchenham B, Charters S. Guidelines for performing systematic literature reviews in software engineering. Technical Report, Version 2.3 EBSE-2007-01, Keele University & University of Durham; 2007.
14. Li R, Liang P, Soliman M, Avgeriou P. Understanding architecture erosion: the practitioners' perceptive. In: Proceedings of the 29th IEEE/ACM International Conference on Program Comprehension (ICPC). IEEE; 2021; Madrid, Spain:311-322.
15. Behnamghader P, Le DM, Garcia J, Link D, Shahbazian A, Medvidovic N. A large-scale study of architectural evolution in open-source software systems. *Empir Softw Eng*. 2017;22(3):1146-1193.
16. Izurieta C, Bieman JM. How software designs decay: a pilot study of pattern evolution. In: Proceedings of the 1st International Symposium on Empirical Software Engineering and Measurement (ESEM). IEEE; 2007:449-451.
17. van Gurp J, Bosch J. Design erosion: problems and causes. *J Syst Softw*. 2002;61(2):105-119.
18. Bandi A, Williams BJ, Allen EB. Empirical evidence of code decay: a systematic mapping study. In: Proceedings of the 20th Working Conference on Reverse Engineering (WCRE). IEEE; 2013; Koblenz, Germany:341-350.
19. Tran JB, Holt RC. Forward and reverse repair of software architecture. In: Proceedings of the Conference of the Centre for Advanced Studies on Collaborative Research (CASCON). IBM Press; 1999; Mississauga, Ontario, Canada:1-9.
20. Parnas DL. Software aging. In: Proceedings of the 16th International Conference on Software Engineering (ICSE). IEEE; 1994; Sorrento, Italy:279-287.
21. Zhang L, Sun Y, Song H, Chauvel F, Mei H. Detecting architecture erosion by design decision of architectural pattern. In: Proceedings of the 23rd International Conference on Software Engineering and Knowledge Engineering (SEKE). KSI; 2011; Miami Beach, FL, USA:758-763.
22. Reimanis D, Izurieta C. Behavioral evolution of design patterns: understanding software reuse through the evolution of pattern behavior. In: Proceedings of the 18th International Conference on Software and Systems Reuse (ICSR). Springer; 2019; Cincinnati, OH, USA:77-93.
23. Hochstein L, Lindvall M. Combating architectural degeneration: a survey. *Informd Softw Technol*. 2005;47(10):643-656.
24. de Oliveira Barros M, de Almeida Farzat F, Travassos GH. Learning from optimization: a case study with Apache Ant. *Inform Softw Technol*. 2015;57:684-704.
25. Basili VR, Caldiera G, Rombach HD. The goal question metric approach. *Encyclopedia Softw Eng*. 1994;1:528-532.
26. Gurgel A, Macia I, Garcia A, et al. Blending and reusing rules for architectural degradation prevention. In: Proceedings of the 13th International Conference on Modularity (Modularity). ACM; 2014; Lugano, Switzerland:61-72.
27. Shaw M, Clements P. The golden age of software architecture. *IEEE Softw*. 2006;23(2):31-39.
28. Kiremire AR. The application of the pareto principle in software engineering. *Consulted Jan*. 2011;13:1-12.
29. Chaniotaki A-M, Sharma T. Architecture smells and pareto principle: a preliminary empirical exploration. In: Proceedings of the 18th IEEE/ACM International Conference on Mining Software Repositories (MSR). IEEE; 2021; Madrid, Spain:190-194.
30. Cohen J. A coefficient of agreement for nominal scales. *Educ Psychol Measurement*. 1960;20(1):37-46.
31. Wohlin C. Second-generation systematic literature studies using snowballing. In: Proceedings of the 20th International Conference on Evaluation and Assessment in Software Engineering (EASE). ACM; 2016; Limerick, Ireland:1-6.
32. Chen L, Babar MA, Zhang H. Towards an evidence-based understanding of electronic data sources. In: Proceedings of the 14th International Conference on Evaluation and Assessment in Software Engineering (EASE). BCS; 2010; Keele University, UK:1-4.
33. Shahin M, Liang P, Babar MA. A systematic review of software architecture visualization techniques. *J Syst Softw*. 2014;94:161-185.
34. Li Z, Avgeriou P, Liang P. A systematic mapping study on technical debt and its management. *J Syst Softw*. 2015;101:193-220.
35. Li R, Liang P, Soliman M, Avgeriou P. Replication package for the paper: understanding software architecture erosion: a systematic mapping study. https://doi.org/10.5281/zenodo.5562418; 2021.
36. Adolph S, Hall W, Kruchten P. Using grounded theory to study the experience of software development. *Empir Softw Eng*. 2011;16(4):487-513.
37. Stol K-J, Ralph P, Fitzgerald B. Grounded theory in software engineering research: a critical review and guidelines. In: Proceedings of the 38th International Conference on Software Engineering (ICSE). ACM; 2016; Austin, TX, USA:120-131.
38. ISO/IEC 25010: 2011 systems and software engineering–systems and software quality requirements and evaluation (SQuaRe)-system and software quality models; 2011.
39. Bass L, Clements P, Kazman R. *Software Architecture in Practice*. 3rd ed.: Addison-Wesley Professional; 2012.
40. Garcia J, Popescu D, Edwards G, Medvidovic N. Identifying architectural bad smells. In: Proceedings of the 13th European Conference on Software Maintenance and Reengineering (CSMR). IEEE; 2009; Kaiserslautern, Germany:255-258.
41. Martin RC. Design principles and design patterns. *Object Mentor*. 2000;1(34):597.
42. Fellah A, Bandi A. On architectural decay prediction in real-time software systems. In: Proceedings of 28th International Conference:98-108.
43. Ma Z, Li R, Li T, et al. A data-driven risk measurement model of software developer turnover. *Soft Comput*. 2020;24(2):825-842.
44. Barney S, Petersen K, Svahnberg M, Aurum A, Barney H. Software quality trade-offs: a systematic map. *Inform Softw Technol*. 2012;54(7):651-662.
45. Conway ME. How do committees invent. *Datamation*. 1968;14(4):28-31.
46. Sturtevant D. Modular architectures make you agile in the long run. *IEEE Softw*. 2017;35(1):104-108.
47. Garlan D, Allen R, Ockerbloom J. Architectural mismatch: why reuse is still so hard. *IEEE Softw*. 2009;26(4):66-69.
48. Grottke M, Matias R, Trivedi KS. The fundamentals of software aging. In: Proceedings of the 19th IEEE International Conference on Software Reliability Engineering Workshops (ISSREW). IEEE; 2008; Seattle, WA, USA:1-6.

49. Ali N, Baker S, O'Crowley R, Herold S, Buckley J. Architecture consistency: state of the practice, challenges and requirements. *Empir Softw Eng*. 2018; 23(1):224-258.
50. Murphy GC, Notkin D, Sullivan KJ. Software reflexion models: bridging the gap between design and implementation. *IEEE Trans Softw Eng*. 2001;27(4): 364-380.
51. Oreizy P, Gorlick MM, Taylor RN, et al. An architecture-based approach to self-adaptive software. *IEEE Intell Syst Appl*. 1999;14(3):54-62.
52. Jaktman CB, Leaney J, Liu M. Structural analysis of the software architecture—a maintenance assessment case study. In: Working Conference on Software Architecture. Springer; 1999:455-470.
53. Wang T, Wang D, Li B. A multilevel analysis method for architecture erosion. In: Proceedings of the 31st International Conference on Software Engineering and Knowledge Engineering (SEKE). KSI; 2019; Hotel Tivoli, Lisbon, Portugal:443-566.
54. Oizumi W, Garcia A, Sousa LDS, Cafeo B, Zhao Y. Code anomalies flock together: exploring code anomaly agglomerations for locating design problems. In: Proceedings of the 38th International Conference on Software Engineering (ICSE). ACM; 2016:440-451.
55. Oizumi WN, Garcia AF, Colanzi TE, Ferreira M, Staa AV. On the relationship of code-anomaly agglomerations and architectural problems. *J Softw Eng Res Develop*. 2015;3(1):1-22.
56. Chen L. Microservices: architecting for continuous delivery and DevOps. In: Proceedings of the 15th IEEE International Conference on Software Architecture (ICSA). IEEE; 2018:39-397.
57. Mumtaz H, Singh P, Blincoe K. A systematic mapping study on architectural smells detection. *J Syst Softw*. 2021;173:110885.
58. Schmitt Laser M, Medvidovic N, Le DM, Garcia J. Arcade: an extensible workbench for architecture recovery, change, and decay evaluation. In: Proceedings of the 28th ACM Joint Meeting on European Software Engineering Conference and Symposium on the Foundations of sofztware Engineering (ESEC/FSE). ACM; 2020; Sacramento, CA, USA:1546-1550.
59. Garcia J, Kouroshfar E, Ghorbani N, Malek S. Forecasting architectural decay from evolutionary history. *IEEE Trans Softw Eng*. 2021;1(1):1-17.
60. Merkle B. Stop the software architecture erosion: building better software systems. In: Companion to the 25th Annual ACM SIGPLAN Conference on Object-Oriented Programming, Systems, Languages, and Applications (SPLASH/OOPSLA). ACM; 2010; Reno/Tahoe, Nevada, USA:129-138.
61. Godfrey MW, Lee EricHS. Secrets from the monster: extracting Mozilla's software architecture. In: Proceedings of the 2nd International Symposium on Constructing Software Engineering Tools (COSET). Citeseer; 2000:1-9.
62. Li Z, Qi X, Yu Q, Liang P, Mo R, Yang C. Multi-programming-language commits in OSS: an empirical study on apache projects. In: Proceedings of the 29th IEEE/ACM International Conference on Program Comprehension (ICPC). IEEE; 2021; Madrid, Spain:219-229.
63. Balalaie A, Heydarnoori A, Jamshidi P. Microservices architecture enables DevOps: migration to a cloud-native architecture. *IEEE Softw*. 2016;33(3): 42-52.
64. Li R, Soliman M, Liang P, Avgeriou P. Symptoms of architecture erosion in code reviews: a study of two OpenStack projects. In: Proceedings of the 19th International Conference on Software Architecture (ICSA). IEEE; 2022; Honolulu, Hawaii, USA.
65. Zhou X, Jin Y, Zhang H, Li S, Huang X. A map of threats to validity of systematic literature reviews in software engineering. In: Proceedings of the 23rd Asia-Pacific Software Engineering Conference (APSEC). IEEE; 2016:153-160.
66. Wohlin C, Runeson P, Höst M, Ohlsson MC, Regnell B, Wesslén A. *Experimentation in Software Engineering*: Springer Science & Business Media; 2012.
67. Baabad A, Zulzalil HB, Baharom SB. Software architecture degradation in open source software: a systematic literature review. *IEEE Access*. 2020;8: 173681-173709.
68. Neri D, Soldani J, Zimmermann O, Brogi A. Design principles, architectural smells and refactorings for microservices: a multivocal review. *SICS Softw-Intensive Cyber-Phys Syst*. 2020;35(1):3-15.
69. Besker T, Martini A, Bosch J. Managing architectural technical debt: a unified model and systematic literature review. *J Syst Softw*. 2018;135:1-16.
70. Herold S, Blom M, Buckley J. Evidence in architecture degradation and consistency checking research: preliminary results from a literature review. In: Proceedings of the 10th European Conference on Software Architecture Workshops (ECSAW). ACM; 2016:1-7.

## APPENDIX A: SELECTED STUDIES

[S1] Li Z & Long J. A case study of measuring degeneration of software architectures from a defect perspective. In: Proceedings of the 18th Asia-Pacific Software Engineering Conference (APSEC). IEEE; 2011; Ho Chi Minh, Vietnam:242-249.

[S2] Baumeister H, Hacklinger F, Hennicker R, Knapp A, Wirsing M. A component model for architectural programming. *Electron Notes Theoret Comput Sci*. 2006;160:75-96.

[S3] Zhao M, Yang J. A DCA-based method for software prognostics and health management. In: Proceedings of the IEEE Prognostics and System Health Management Conference (PHM). IEEE; 2012; Beijing, China:1-5.

[S4] Terra R, Valente MT. A dependency constraint language to manage object-oriented software architectures. *Softw Pract Exper*. 2009;39(12):1073-1094.

[S5] Rama GM. A desiderata for refactoring-based software modularity improvement. In: Proceeding of the 3rd Annual India Software Engineering Conference (ISEC). ACM; 2010; Mysore, India:93-102.

[S6] Mokni A, Urtado C, Vauttier S, Huchard M, Zhang HY. A formal approach for managing component-based architecture evolution. *Sci Comput Programm*. 2016;127:24-49.

[S7] Ayyaz S, Rehman S, Qamar U. A four method framework for fighting software architecture erosion. *Int J Comput Control, Quantum Inform Eng*. 2015;9(1):133-139.

[S8] Bouwers E, van Deursen A. A lightweight sanity check for implemented architectures. *IEEE Softw*. 2010;27(4):44-50.

[S9] Izurieta C, Bieman JM. A multiple case study of design pattern decay, grime, and rot in evolving software systems. *Softw Qual J*. 2013;21(2):289-323.

[S10] Mirakhorli M, Cleland-Huang J. A pattern system for tracing architectural concerns. In: Proceedings of the 18th Conference on Pattern Languages of Programs (PLoP). ACM; 2011; Portland, Oregon, USA:1-10.

[S11] Sejfia A. A pilot study on architecture and vulnerabilities: lessons learned. In: Proceedings of the 2nd IEEE/ACM International Workshop on Establishing the Community-Wide Infrastructure for Architecture-Based Software Engineering (ECASE). IEEE; 2019; Montreal, Quebec, Canada:42-47.

[S12] Terra R, Valente MT, Czarnecki K, Bigonha RS. A recommendation system for repairing violations detected by static architecture conformance checking. *Softw Pract Exper*. 2015;45(3):315-342.

[S13] Herold S, Rausch A. A rule-based approach to architecture conformance checking as a quality management measure. *Relating System Quality and Software Architecture*: Morgan Kaufmann; 2014:181-207.

[S14] Caracciolo A, Lungu MF, Nierstrasz O. A unified approach to architecture conformance checking. In: Proceedings of the 12th Working IEEE/IFIP Conference on Software Architecture (WICSA). IEEE; 2015; Montreal, QC, Canada:41-50.

[S15] Hassaine S, Guéhéneuc Y-G, Hamel S, Antoniol G. ADvISE: architectural decay in software evolution. In: Proceedings of the 16th European Conference on Software Maintenance and Reengineering (CSMR). IEEE; 2012; Szeged, Hungary:267-276.

[S16] Le DM, Behnamghader P, Garcia J, Link D, Shahbazian A, Medvidovic N. An empirical study of architectural change in open-source software systems. In: Proceedings of the 12th IEEE/ACM Working Conference on Mining Software Repositories (MSR). IEEE; 2015; Florence, Italy:235-245.

[S17] Le DM, Link D, Shahbazian A, Medvidovic N. An empirical study of architectural decay in open-source software. In: Proceedings of the 15th IEEE International Conference on Software Architecture (ICSA). IEEE; 2018; Seattle, WA, USA:176-185.

[S18] Fontana FA, Roveda R, Zanoni M, Raibulet C, Capilla R. An experience report on detecting and repairing software architecture erosion. In: Proceedings of the 13th Working IEEE/IFIP Conference on Software Architecture (WICSA). IEEE; 2016; Venice, Italy:21-30.

[S19] Mirakhorli M, Fakhry A, Grechko A, Wieloch M, Cleland-Huang J. Archie: a tool for detecting, monitoring, and preserving architecturally significant code. In: Proceedings of the 22nd ACM/SIGSOFT International Symposium on Foundations of Software Engineering (FSE). ACM; 2014; Hong Kong, China:739-742.

[S20] Greifenberg T, Müller K, Rumpe B. Architectural consistency checking in plugin-based software systems. In: Proceedings of the 2nd Workshop on Software Architecture Erosion and Architectural Consistency (SAEroCon). ACM; 2015; Dubrovnik/Cavtat, Croatia:1-7.

[S21] Riaz M, Sulayman M, Naqvi H. Architectural decay during continuous software evolution and impact of 'design for change' on software architecture. In: Proceedings of the International Conference on Advanced Software Engineering and Its Applications (ASEA). Springer; 2009; Jeju Island, Korea:119-126.

[S22] Bhattacharya S, Perry DE. Architecture assessment model for system evolution. In: Proceedings of the 6th Working IEEE/IFIP Conference on Software Architecture (WICSA). IEEE; 2007; Mumbai, Maharashtra, India:44-53.

[S23] Pruijt L, Köppe C, Brinkkemper S. Architecture compliance checking of semantically rich modular architectures: a comparative study of tool support. In: Proceedings of the 29th IEEE International Conference on Software Maintenance (ICSM). IEEE; 2013; Eindhoven, The Netherlands:220-229.

[S24] Miranda S, Rodrigues JrE, Valente MT, Terra R. Architecture conformance checking in dynamically typed languages. *J Object Technol*. 2016;15(3):1-34.

[S25] Schröder S, Soliman M, Riebisch M. Architecture enforcement concerns and activities—an expert study. *J Syst Softw*. 2018;145:79-97.

[S26] Guimarães E, Garcia A, Cai Y. Architecture-sensitive heuristics for prioritizing critical code anomalies. In: Proceedings of the 14th International Conference on Modularity (MODULARITY). ACM; 2015; Fort Collins, CO, USA:68-80.

[S27] Macia I, Garcia J, Popescu D, Garcia A, Medvidovic N, von Staa A. Are automatically-detected code anomalies relevant to architectural modularity?: An exploratory analysis of evolving systems. In: Proceedings of the 11th Annual International Conference on Aspect-oriented Software Development (AOSD). ACM; 2012; Potsdam, Germany:167-178.

[S28] Bandi A, Allen EB, Williams BJ. Assessing code decay: a data-driven approach. In: Proceedings of the 24th International Conference on Software Engineering and Data Engineering (SEDE). ISCA; 2015; San Diego, CA, USA:1-8.

[S29] Neto VVG, Manzano W, Garcés L, Guessi M, Oliveira B, Volpato T, Nakagawa EY. Back-SoS: towards a model-based approach to address architectural drift in systems-of-systems. In: Proceedings of the 33rd Annual ACM Symposium on Applied Computing (SAC). ACM; 2018; Pau, France:1461-1463.

[S30] Reimanis D, Izurieta C. Behavioral evolution of design patterns: understanding software reuse through the evolution of pattern behavior. In: Proceedings of the 18th International Conference on Software and Systems Reuse (ICSR). Springer; 2019; Cincinnati, OH, USA:77-93.

[S31] Gurgel A, Macia I, Garcia A, et al. Blending and reusing rules for architectural degradation prevention. In: Proceedings of the 13th International Conference on Modularity (MODULARITY). ACM; 2014; Lugano, Switzerland:61-72.

[S32] Langhammer M. Co-evolution of component-based architecture-model and object-oriented source code. In: Proceedings of the 18th International Doctoral Symposium on Components and Architecture (WCOP). ACM; 2013; Vancouver, BC, Canada:37-42.

[S33] Herold S, Rausch A. Complementing model-driven development for the detection of software architecture erosion. In: Proceedings of the 5th International Workshop on Modeling in Software Engineering (MiSE). IEEE; 2013; San Francisco, CA, USA:24-30.

[S34] De Silva L, Balasubramaniam D. Controlling software architecture erosion: a survey. *J Syst Softw*. 2012;85(1):132-151.

[S35] Rocha H, Durelli RS, Terra R, Bessa S, Valente MT. DCL 2.0: modular and reusable specification of architectural constraints. *J Brazilian Comput Soc*. 2017;23(1):1-25.

[S36] Bertran IM. Detecting architecturally-relevant code smells in evolving software systems. In: Proceedings of the 33rd International Conference on Software Engineering (ICSE). ACM; 2011; Waikiki, Honolulu, HI, USA:1090-1093.

[S37] Zhang L, Sun Y, Song H, Chauvel F, Mei H. Detecting architecture erosion by design decision of architectural pattern. In: Proceedings of the 23rd International Conference on Software Engineering and Knowledge Engineering (SEKE). KSI; 2011; Miami Beach, FL, USA:758-763.

[S38] Wong S, Cai Y, Kim M, Dalton M. Detecting software modularity violations. In: Proceedings of the 33rd International Conference on Software Engineering (ICSE). ACM; 2011; Waikiki, Honolulu, HI, USA:411-420.

[S39] Mirakhorli M, Cleland-Huang J. Detecting, tracing, and monitoring architectural tactics in code. *IEEE Trans Softw Eng*. 2016;42(3):206-221.

[S40] Herold S, English M, Buckley J, Counsell S, Cinnéide MÓ. Detection of violation causes in reflexion models. In: Proceedings of the 22nd IEEE International Conference on Software Analysis, Evolution, and Reengineering (SANER). IEEE; 2015; Montreal, QC, Canada:565-569.

[S41] Pérez-Castillo R, de Guzmán IGR, Piattini M. Diagnosis of software erosion through fuzzy logic. In: Proceedings of the IEEE Symposium on Computational Intelligence in Dynamic and Uncertain Environments (CIDUE). IEEE; 2011; Paris, France:49-56.

[S42] Nam D, Lee YK, Medvidovic N. Eva: A tool for visualizing software architectural evolution. In: Proceedings of the 40th International Conference on Software Engineering (ICSE) Companion. ACM; 2018; Gothenburg, Sweden:53-56.

[S43] Altınışık M, Ersoy E, Sözer H. Evaluating software architecture erosion for PL/SQL programs. In: Proceedings of the 11th European Conference on Software Architecture (ECSA) Companion. ACM; 2017; Canterbury, United Kingdom:159-165.

[S44] Lindvall M, Becker M, Tenev V, Duszynski S, Hinchey M. Good change and bad change: an analysis perspective on software evolution. *Trans Found Master Change I*. 2016;9960:90-112.

[S45] Bandara V, Perera I. Identifying software architecture erosion through code comments. In: Proceedings of the 18th International Conference on Advances in ICT for Emerging Regions (ICTer). IEEE; 2018; Colombo, Sri Lanka:62-69.

[S46] Chhabra JK. Improving package structure of object-oriented software using multi-objective optimization and weighted class connections. *J King Saud Univ -Comput Inform Sci*. 2017;29(3):349-364.

[S47] de Oliveira Barros M, de Almeida Farzat F, Travassos GH. Learning from optimization: a case study with Apache Ant. *Inform Softw Technol*. 2015;57:684-704.

[S48] Dimech C, Balasubramaniam D. Maintaining architectural conformance during software development: a practical approach. In: Proceedings of the 7th European Conference on Software Architecture (ECSA). Springer; 2013; Montpellier, France:208-223.

[S49] Mo R, Garcia J, Cai Y, Medvidovic N. Mapping architectural decay instances to dependency models. In: Proceedings of the 4th International Workshop on Managing Technical Debt (MTD). IEEE; 2013; San Francisco, CA, USA:39-46.

[S50] Strasser A, Cool B, Gernert C, et al. Mastering erosion of software architecture in automotive software product lines. In: Proceedings of the 40th International Conference on Current Trends in Theory and Practice of Informatics (SOFSEM). Springer; 2014; Smokovec, Slovakia:491-502.

[S51] Koziolek H, Domis D, Goldschmidt T, Vorst P. Measuring architecture sustainability. *IEEE Softw*. 2013;30(6):54-62.

[S52] Koziolek H, Domis D, Goldschmidt T, Vorst P, Weiss RJ. MORPHOSIS: a lightweight method facilitating sustainable software architectures. In: Proceedings of the Joint Working IEEE/IFIP Conference on Software Architecture and European Conference on Software Architecture (WICSA/ECSA). IEEE; 2012; Helsinki, Finland:253-257.

[S53] Olsson T, Ericsson M, Wingkvist A. Motivation and impact of modeling erosion using static architecture conformance checking. In: Proceedings of the 1st IEEE International Conference on Software Architecture Workshops (ICSAW). IEEE; 2017; Gothenburg, Sweden:204-209.

[S54] Schmidt F, MacDonell S, Connor AM. Multi-objective reconstruction of software architecture. *Int J Softw Eng Knowl Eng*. 2018;28(6):869-892.

[S55] Brunet J, Bittencourt RA, Serey D, Figueiredo J. On the evolutionary nature of architectural violations. In: Proceedings of the 19th Working Conference on Reverse Engineering (WCRE). IEEE; 2012; Kingston, ON, Canada:257-266.

[S56] Jaafar F, Hassaine S, Guéhéneuc Y-G, Hamel S, Adams B. On the relationship between program evolution and fault-proneness: an empirical study. In: Proceedings of the 17th European Conference on Software Maintenance and Reengineering (CSMR). IEEE; 2013; Genova, Italy:15-24.

[S57] Macia I, Arcoverde R, Garcia A, Chavez C, von Staa A. On the relevance of code anomalies for identifying architecture degradation symptoms. In: Proceedings of the 16th European Conference on Software Maintenance and Reengineering (CSMR). IEEE; 2012; Szeged, Hungary:277-286.

[S58] Juarez Filho L, Rocha L, Andrade R, Britto R. Preventing erosion in exception handling design using static-architecture conformance checking. In: Proceedings of the 11th European Conference on Software Architecture (ECSA). Springer; 2017; Canterbury, UK:67-83.

[S59] Bakota T, Hegedűs P, Siket I, Ladányi G, Ferenc R. QualityGate SourceAudit: a tool for assessing the technical quality of software. In: Proceedings of the IEEE Conference on Software Maintenance, Reengineering, and Reverse Engineering (CSMR-WCRE). IEEE; 2014; Antwerp, Belgium:440-445.

[S60] Herold S, Mair M. Recommending refactorings to re-establish architectural consistency. In: Proceedings of the 8th European Conference on Software Architecture (ECSA). Springer; 2014; Vienna, Austria:390-397.

[S61] Haitzer T, Navarro E, Zdun U. Reconciling software architecture and source code in support of software evolution. *J Syst Softw*. 2017;123:119-144.

[S62] Stal M. Refactoring software architectures. *Agile Software Architecture, Chapter 3*: Elsevier; 2014:63-82.

[S63] Adersberger J, Philippsen M. ReflexML: UML-based architecture-to-code traceability and consistency checking. In: Proceedings of the 5th European Conference on Software Architecture (ECSA). Springer; 2011; Essen, Germany:344-359.

[S64] Le DM, Carrillo C, Capilla R, Medvidovic N. Relating architectural decay and sustainability of software systems. In: Proceedings of the 13th Working IEEE/IFIP Conference on Software Architecture (WICSA). IEEE; 2016; Venice, Italy:178-181.

[S65] Passos L, Terra R, Valente MT, Diniz R, Mendonça N. Static architecture-conformance checking An illustrative overview. *IEEE Softw*. 2010;27(5):82-89.

[S66] Merkle B. Stop the software architecture erosion: building better software systems. In: Proceedings of the 25th Annual ACM SIGPLAN Conference on Object-Oriented Programming, Systems, Languages, and Applications (SPLASH/OOPSLA) Companion. ACM; 2010; Reno/Tahoe, Nevada, USA:129-138.

[S67] Nicolaescu A, Lichter H, Göinger A, Alexander P, Le D. The ARAMIS workbench for monitoring, analysis and visualization of architectures based on run-time interactions. In: Proceedings of the 2nd Workshop on Software Architecture Erosion and Architectural Consistency (SAEroCon). ACM; 2015; Dubrovnik/Cavtat, Croatia:1-7.

[S68] Feilkas M, Ratiu D, Jurgens E. The loss of architectural knowledge during system evolution: an industrial case study. In: Proceedings of the 17th IEEE International Conference on Program Comprehension (ICPC). IEEE; 2009; Vancouver, BC, Canada:188-197.

[S69] Mair M, Herold S, Rausch A. Towards flexible automated software architecture erosion diagnosis and treatment. In: Proceedings of the 11th Working IEEE/IFIP Conference on Software Architecture (WICSA) Companion. ACM; 2014; Sydney, NSW, Australia:1-6.

[S70] Gerdes S, Jasser S, Riebisch M, Schröder S, Soliman M, Stehle T. Towards the essentials of architecture documentation for avoiding architecture erosion. In: Proceedings of the Workshop on Sustainable Architecture: Global Collaboration, Requirements, Analysis (SAGRA). ACM; 2016; Copenhagen, Denmark:1-4.

[S71] Medvidovic N, Jakobac V. Using software evolution to focus architectural recovery. *Automated Softw Eng*. 2006;13(2):225-256.

[S72] Baum D, Dietrich J, Anslow C, Müller R. Visualizing design erosion: how big balls of mud are made. In: Proceedings of the 6th IEEE Working Conference on Software Visualization (VISSOFT). IEEE; 2018; Madrid, Spain:122-126.

[S73] Dalgarno M. When good architecture goes bad. *Methods and Tools*. 2009;17(1):27-34.

## APPENDIX B: ABBREVIATIONS USED IN THIS SMS

| | |
|---|---|
| ACC | Architecture conformance checking |
| ADL | Architecture description language |
| ADvISE | Architectural decay in software evolution |
| AEr | Architecture erosion |
| ARAMIS | Architecture analysis and monitoring infrastructure |
| DCA | Discriminant coordinates analysis |
| DCL | Dependency constraint language |
| DSL | Domain-specific language |
| DSM | Design structure matrix |
| FIS | Fuzzy inference system |
| GQM | Goal–question–metric |
| LiSCIA | Light-weight sanity check for implemented architectures |
| MSA | Microservices architecture |
| PHM | Prognostics and health management |
| QA | Quality attribute |
| RM | Reflexion modeling |
| RQ | Research question |
| SLR | Systematic literature review |
| SMS | Systematic mapping study |
| SoS | System-of-systems |
| VA | Variant analysis |